\documentclass[preprint,5p,times,twocolumn]{elsarticle}
\usepackage{amssymb}
\usepackage{graphicx}
\usepackage{amsmath}
\usepackage{amsthm}
\usepackage[usenames,dvipsnames]{color}
\biboptions{numbers,square,comma,sort&compress}

\theoremstyle{plain}

\theoremstyle{definition}
\newtheorem{definition}{Definition}[section]
\theoremstyle{remark}

\providecommand{\mb}[1]{\mathbf{#1}}

\newcommand{\captionfonts}{\small}
\makeatletter  
\long\def\@makecaption#1#2{%
  \vskip\abovecaptionskip
  \sbox\@tempboxa{{\captionfonts #1: #2}}%
  \ifdim \wd\@tempboxa >\hsize
    {\captionfonts #1: #2\par}
  \else
    \hbox to\hsize{\hfil\box\@tempboxa\hfil}%
  \fi
  \vskip\belowcaptionskip}
\makeatother   

\newcommand{\bibpath}{./}

\newcommand{\comment}[1]{}

\journal{Mathematical Biosciences}
\begin{document}
\begin{frontmatter}

\title{Determining Structurally Identifiable Parameter Combinations Using Subset Profiling}
\author[UM,UMmath]{Marisa C. Eisenberg\corref{cor1}}
\ead{marisae@umich.edu}
\author[UM]{Michael A. L. Hayashi\corref{cor1}}
\ead{mhayash@umich.edu}

\address[UM]{Department of Epidemiology, School of Public Health, University of Michigan, Ann Arbor}
\address[UMmath]{Department of Mathematics, University of Michigan, Ann Arbor}
\cortext[cor1]{Corresponding author.}




\begin{abstract} 
Identifiability is a necessary condition for successful parameter estimation of dynamic system models. A major component of identifiability analysis is determining the identifiable parameter combinations, the functional forms for the dependencies between unidentifiable parameters. Identifiable combinations can help in model reparameterization and also in determining which parameters may be experimentally measured to recover model identifiability. Several numerical approaches to determining identifiability of differential equation models have been developed, however the question of determining identifiable combinations remains incompletely addressed. In this paper, we present a new approach which uses parameter subset selection methods based on the Fisher Information Matrix, together with the profile likelihood, to effectively estimate identifiable combinations. We demonstrate this approach on several example models in pharmacokinetics, cellular biology, and physiology. 
\\
\end{abstract}


\end{frontmatter}

\section{Introduction}  \label{sec:intro}
Identifiability analysis is a critical step in the parameter estimation process which addresses whether it is possible to uniquely recover the model parameters from a given set of data. For ordinary differential equation (ODE) models, this problem is often broken into two broad and often overlapping categories: \emph{practical} or \emph{numerical identifiability}, which incorporates practical estimation issues that come with real data (such as noise and bias), and \emph{structural identifiability}, which considers a best-case scenario when the data are assumed to be known completely (i.e. smooth, noise-free and known for every time point).  Structural identifiability is a necessary condition for parameter estimation with noisy data \cite{Cobelli1980, Meshkat2009}.  

In the common case of model unidentifiability, a key concept in identifiability analysis is that of \emph{identifiable combinations}, i.e. combinations of parameters which are identifiable even if the individual parameters are not \cite{Cobelli1980, Meshkat2009}. These combinations give information on how to reparameterize a given model for identifiability and also give insight into what additional parameters can be experimentally measured to yield an identifiable model \cite{Meshkat2009}.  

Many different analytical approaches to structural identifiability have been developed \cite{Bellu2007, Chappell1998, Cobelli1980, Pohjanpalo1978, Vajda1989}.  However, these methods are often restricted to specific classes of models (e.g. Laplace transforms \cite{Bellman1970, Cobelli1980, DiStefano1983}, differential algebra \cite{Audoly2001, Meshkat2009}). They may also be difficult to implement algorithmically, computationally intensive, or not guaranteed to terminate, making applications beyond relatively simple models more challenging \cite{Saccomani2003, Vajda1989, Pohjanpalo1978, Chis2011}.  One common method for identifiability of polynomial and rational function ODE models is via differential algebra \cite{Bellu2007, Ljung1994, Saccomani2003}. In the case of unidentifiability, the differential algebra approach can also be used to uncover identifiable parameter combinations and reparameterizations of the model in terms of these combinations \cite{Meshkat2009}. The differential algebra-based method in \cite{Meshkat2009} uses Gr\"{o}bner bases to find a `simplest' set of combinations, denoted the canonical set. However, this can require the expensive calculation of large numbers of Gr\"{o}bner bases \cite{Meshkat2009}. 

By contrast, while most numerical approaches to identifiability provide only local (rather than global) information about the parameters, they are often more computationally tractable \cite{Hengl2007}. In some cases these methods can be used to address both structural and practical identifiability, e.g. by using simulated data without errors (so that any resulting unidentifiability must be due to structural identifiability issues) \cite{Raue2009, Eisenberg2008, Eisenberg2013}. Techniques for determining model identifiability include the Fisher Information Matrix (FIM) \cite{Jacquez1985,Jacquez1990, CintronArias2009, Botelho2012, Komorowski2011} and the profile likelihood \cite{Raue2009}, among others \cite{Raue2013,BalsaCanto2010, Viallefont1998}. However, the problem of finding identifiable combinations for nonlinear ODE models has received less attention \cite{Raue2009, Hengl2007}. 

In this paper, we propose a simple numerical approach to determining structurally identifiable combinations. We make use of two established tools in numerical identifiability analysis: the FIM (and associated Cramer-Rao estimates of the covariance matrix), and the profile likelihood \cite{Jacquez1985, Jacquez1990, CintronArias2009, Raue2009}. Individually, there are gaps in the applicability of each method (particularly for higher dimensional combinations) \cite{Raue2009}, as discussed further below. Instead, our approach builds on these two tools to determine structurally identifiable combinations in nonlinear differential equation models.

\section{Framework and Definitions}  \label{sec:framework}
\subsection{Model Structure}
We begin by introducing the overall modeling framework and identifiability definitions used here. Let the model be given by
\begin{equation}
\begin{aligned}
\dot{\mb{x}} &= f(\mb{x},t,\mb{u,p})\\ 
\mb{y} &= g(\mb{x},t,\mb{p})
\end{aligned}
\label{eq:modelsetup}
\end{equation}
where $\dot{\mb{x}}$ is a system of first order ODEs, with $t$ representing time, and $\mb u$ the experimental input function(s), if any.  The set of model parameters are given by $\mathbf{p}$ (typically real-valued). 
The model output(s) are given by $\mb{y}$, which represents the measured variables---in our case assumed to be noise-free.  We also let $\bf x_0$ represent the vector of initial conditions for $\mb x (t)$. 

\subsection{Identifiability}
Identifiability analysis explores the question: given an input $\mb{u}$, model $\dot{\mb{x}} = f(\mb{x},t, \mb{u,p})$ and experimental output $\mb{y}$, is it possible to uniquely identify the parameters $\mb{p}$? Structural identifiability examines a `best-case' version of this question in which we assume `perfect' noiseless data. If parameter has a unique value $p^*$ which yields a given output $\mb y^*$, it is considered \emph{globally (or uniquely) structurally identifiable}; if there is a unique value $p^*$ within a local neighborhood of parameter space yielding $\mb y^*$, it is considered \emph{locally structurally identifiable}; and if there are a continuum of values of $p$ which yield the output $\mb y^*$, the parameter is considered \emph{unidentifiable}. A model is said to be globally structurally identifiable if all the parameters are globally structurally identifiable; if any parameters are locally structurally identifiable or unidentifiable, the model is also considered locally structurally identifiable or unidentifiable, respectively. In the case of model unidentifiability, the model parameters typically form \emph{identifiable combinations}, i.e. combinations of parameters which are identifiable even though the individual parameters are unidentifiable. 

More formally, structural identifiability can be thought of in terms of injectivity of the map $\Phi: \bf p \rightarrow y$ given by viewing the model output $\mb{y}$ as a function of the parameters $\mb{p}$ \cite{Meshkat2009, Saccomani2003}. We note that because there may be some `special' or degenerate parameter values or initial conditions for which an otherwise identifiable model is unidentifiable (e.g. if all initial conditions are zero and there is no input to the model), structural identifiability is often defined for almost all parameter values and initial conditions \cite{Meshkat2009, Audoly2001, Saccomani2003}. 

\begin{figure}
\centering
\includegraphics[width=0.33\textwidth]{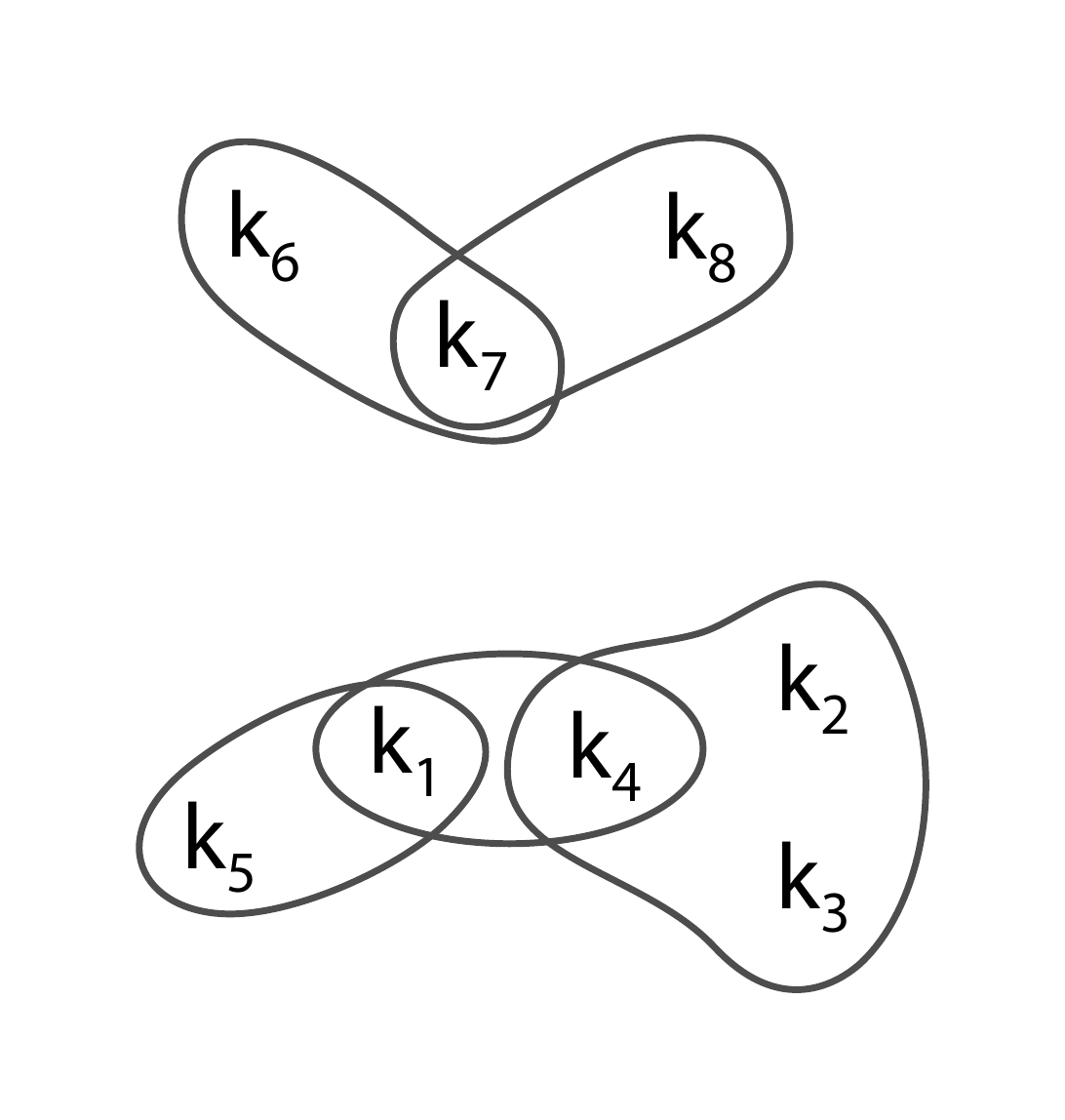}
\caption{An example set of parameter combinations. Circles indicate the five identifiable parameter combinations, e.g. $k_1$ and $k_5$ are involved in an identifiable combination. Some parameters are involved in more than one identifiable combination (e.g. $k_1$), so that there are two overall connected components.}
\label{fig:looseparams}
\end{figure}

\begin{definition} For a given ODE model $\dot{\mb{x}} = f(\mb{x},t,\mb{u,p})$ and output  $\mb{y}$, an individual parameter $p$ is \emph{uniquely (or globally) structurally identifiable} if for almost every point $\mb{p}^*$ and almost all initial conditions, the equation $\mb{y}(\mb{x},t,\mb{p}^*) = \mb{y}(\mb{x},t,\mb{p})$ implies $p = p^*$.  A parameter $p$ is said to be \emph{non-uniquely (or locally) structurally identifiable} if for almost any $\mb{p}^*$ and almost all initial conditions, the equation $\mb{y}(\mb{x},t,\mb{p}^*) = \mb{y}(\mb{x},t,\mb{p})$ implies that $p$ has more than one solution, but each solution is unique within a local neighborhood of the parameters. Otherwise, a parameter is said to be \emph{unidentifiable}.  \end{definition}

\begin{definition}  Similarly, a model $\dot{\mb{x}} = f(\mb{x},t,\mb{u,p})$ is said to be \emph{uniquely} (respectively \emph{non-uniquely}) \emph{structurally identifiable}  for a given choice of output $\mb{y}$ if every parameter is uniquely (respectively non-uniquely) structurally identifiable, i.e. the equation $\mb{y}(\mb{x},t,\mb{p}^*) = \mb{y}(\mb{x},t,\mb{p})$ has only one solution, $\bf p = p^*$ (respectively finitely many solutions).  Equivalently, a model is uniquely structurally identifiable for a given output if and only if the map $\Phi$ is injective almost everywhere, i.e. if there exists a unique set of parameter values $\mb{p}^*$ which yields a given trajectory $\mb{y}(\mb{x},t,\mb{p}^*)$ almost everywhere. \end{definition}

\subsection{Parameter Graph}
In examining the parameter identifiability structure, it is often convenient to consider a parameter graph of the identifiable combinations. We draw this as a hypergraph with the parameters as nodes and the identifiable combinations as edges, with an example shown in Figure \ref{fig:looseparams}. As we will see in Sections \ref{sec:numapproach} and \ref{sec:exs}, the structure and connected components of the parameter graph will be used to precondition the degrees of freedom when estimating parameters in the likelihood profiles. 

\subsection{Fisher Information Matrix}
The Fisher Information Matrix $\mathbf{F}$ is an $N \times N$ symmetric matrix that represents the amount of information contained in the data, $\mathbf{y}^*$, about parameters $\mathcal{A} = \left\{p_1, ..., p_N\right\}$  where $\mathcal{A} \subset \mathbf{p}$ \cite{Jacquez1985, Jacquez1990}.  If $\mathbf{F}$ is singular, $\mathcal{A}$ is unidentifiable.  In practice, $\mathcal{A}$ may be unidentifiable when the determinant of $\mathbf{F}$ is non-zero but small.  Regardless, the rank of the FIM corresponds to the number of identifiable parameters or combinations in $\mathcal{A}$ \cite{Komorowski2011,CintronArias2009, Cobelli1980}.  Inverting the FIM results in the Cram\'{e}r-Rao bound Covariance Matrix, $\mathbf{C}$.  The diagonal entries of $\mathbf{C}$ correspond to the individual variances of parameters in $\mathcal{A}$.  If $\mathcal{A}$ is a singleton parameter set $\{p\}$, the variance $\mathbf{C}$ is simply the reciprocal of the squared parameter sensitivity, $\frac{1}{(dy/dp)^2}$.  Given a model $\dot{\mb{x}}$ and corresponding output function $\mb{y}$, Fisher Information Matrices for parameter sets can be computed as follows:
\begin{enumerate}
	\item Generate the sensitivity matrix 
		\begin{equation} 
			\mathbf{X} = (\mathbf{s}(\mathbf{t}, p_1), ... , \mathbf{s}(\mathbf{t}, p_N))
		\end{equation}
		where $\mathbf{s}(\mathbf{t}, p_i) = [\frac{\delta \mathbf{y}}{\delta p_i}(t_0), ..., \frac{\delta \mathbf{y}}{\delta p_i}(t_t)]^T$.
	\item Compute the Fisher Information Matrix:
		\begin{equation}
			\mathbf{F} = \mathbf{X}^T \mathbf{X}
		\end{equation}
	\item Provided $\mathbf{F}$ is not singular, compute the Covariance Matrix:
		\begin{equation}
			\mathbf{C} = \mathbf{F}^{-1}
		\end{equation}
\end{enumerate}

We will often refer to the rank of the FIM generated from the parameters in a given connected component as the \emph{rank of the component}, or equivalently the number of identifiable combinations in the component. More broadly, we refer to the rank the FIM generated by a particular subset of parameters as the rank of that subset. When the number of parameters in a connected component ($N_p$) is more than one greater than the number of identifiable combinations (i.e. $N_p>N_c+1$, where $N_c$ is the rank of the component), we will denote this as a component containing `loose' parameters, with an example given in Figure \ref{fig:looseparams} (the lower component, due to the pair $k_2$, $k_3$). 

While the FIM and covariance matrix are useful in determining the overall identifiability of the parameters, it is not straightforward to uncover precisely which parameters are involved in a given combination. This is particularly an issue in the common scenario where multiple identifiable combinations share overlapping parameters (e.g. if $p_1p_2$ and $p_1+p_3$ are both identifiable combinations). In this case, while one can detect which parameters are unidentifiable (e.g. by examining their variances in $\mathbf{C}$), the specific parameters involved in each combination is more difficult to determine. Moreover, as the FIM is evaluated at a single point in parameter space, this approach cannot determine the functional form of the combinations. 

\subsection{Profile Likelihood Approach}
Another common tool in assessing parameter identifiability is the profile likelihood \cite{Raue2009,Murphy2000,Venzon1988}.  This approach `profiles' a single parameter $p_i \in \mathcal{A}$ by fixing the value of $p_i$ across a range of values, and fitting all remaining parameters for each fixed value of $p_i$, using the likelihood function $\mathcal{L}$ as the objective function.  The maximum value of the likelihood function for each parameter value constitutes the likelihood profile for the fixed parameter.  A parameter is structurally unidentifiable when its likelihood profile is flat across its range.  Alternatively, a parameter is practically unidentifiable when the curvature of its likelihood profile is shallow.  For unidentifiable parameters, the best fit values of the other parameters may indicate the functional form of pairwise projections of identifiable combinations (as discussed in \cite{Raue2009}).  Given data $\mathbf{y}^*$ and parameters $\mathcal{A}$, the following algorithm computes the likelihood profile for each parameter in $\mathcal{A}$:
\begin{itemize}
	\item Simulate data $\mathbf{y}$ for a starting parameter point $\mathbf{p}^*$.
	\item For each $p_i \in \mathcal{A}$, select a range of values from $[\min{p_i}, \max{p_i}]$, about the best fit value $\hat{p_i}$. 
	\begin{itemize}
		\item For each value of $p_i$ in $[\min{p_i}, \max{p_i}]$:
		 \begin{enumerate}
			\item Estimate the remaining parameters $p_j \in \mathcal{A}\setminus p_i$
			\item Record best-fit values of the parameters and $\mathcal{L}$
			\end{enumerate}
		\end{itemize}
\end{itemize}

In practice, the range $[\min{p_i}, \max{p_i}]$ can be determined dynamically based on a threshold for $\mathcal{L}$ \cite{Raue2009}. Likelihood profiles can in some cases be used to determine the functional form of the identifiable combinations---by examining how the estimates of the remaining parameters change as we profile a particular parameter $p_i$, we can trace out the form of the identifiable combinations (as developed in \cite{Raue2009}). For example, if we have a combination $p_1 + p_2$, then when profiling $p_1$, we would expect $p_2$ to change in a compensatory way which preserves the sum $p_1 + p_2$, as this will preserve the best fit to the data. 

As noted in \cite{Raue2009}, this approach is ill-conditioned when there are multiple parameters in a combination or multiple combinations. Any extra degree(s) of freedom in the fitted parameters allows them to compensate for one another, such that they aren't constrained to trace out the form of the identifiable combinations with the profiled parameter. For example,  when profiling $p_1$, if our combination is $p_1 + p_2 + p_3$ then there are infinitely many ways that $p_2$ and $p_3$ can compensate to maintain the sum $p_1+p_2+p_3$ (and thus maintain the same fit to the data). The resulting profiled parameter relationships are often noisy or arbitrary, as there is a range of values for the unidentifiable parameters which will yield the same output in the profile. This impairs fitting functions to the profile parameter relationships, as illustrated in Example 2 below. These issues can be addressed by restricting the set of parameters used for profiling to maintain appropriate degrees of freedom. To this end, Hengl et al. present the optimal transformation method \cite{Hengl2007}. Here we propose an alternate approach using the FIM.

\section{Numerical Approach to Identifiable Combinations} \label{sec:numapproach}
We begin with the sensitivity matrix for the complete model where $\mathcal{A} = \mathbf{p}$.  To determine which parameters are identifiable, we compute a FIM and its associated covariance matrix.  We identify insensitive parameters by examining individual parameter variances.  Identifiable and insensitive parameters are omitted for the remainder of the analysis.  

A central element of our approach is establishing identifiable combinations in the parameter graph.  We accomplish this task by performing a systematic search over parameter subset ranks (the rank of the FIM for subset $\mathcal{A}_i$).  We say a rank-deficient subset has \emph{nearly full rank} when the exclusion of any parameter yields a full rank FIM.  Subsets which are nearly full rank can be assembled into connected components of parameters. By profiling these subsets, we condition the parameter estimation for each likelihood profile to maintain the appropriate degrees of freedom.  
We can then use the plots of pairwise parameter relationships to construct the functional forms of identifiable combinations.
\pagebreak

\noindent The steps of the method are as follows:
\begin{enumerate}
\item \textbf{Complete Model FIM} - 
We first determine the rank of the FIM for $\mathbf{p}$, which allows us to determine the number of identifiable parameters/combinations. We invert the FIM to generate an overall covariance matrix. The resulting variances of the model parameters allow us to uncover any identifiable parameters, as these need not be considered for determining combinations in the subsequent steps. We use the coefficient of variation (given by $\%CV = 100 \cdot SD/p$, where $p$ is the value of the parameter) as our parameter uncertainty measure, as this accounts for the size of the parameter value when evaluating parameter uncertainty.  We take a tolerance of $\%CV \geq100$ to indicate structural unidentifiability in the parameters (and suppose that parameters may be practically unidentifiable for $\sim25\leq\%CV<100$), although when parameters are structurally unidentifiable, often the $\%CV$'s are much larger, e.g. $>10^6$ (see Examples below). 
\\
\item \textbf{Single Parameter Sensitivity} - Similarly, we can exclude parameters which are \emph{completely insensitive}, i.e. which are unidentifiable even with all the other parameters fixed. These can be identified by examining the sensitivity of the output to each parameter, or equivalently the variances from the FIM generated if we assume we are fitting each parameter individually and fixing the rest (these are simply the inverses of the diagonal elements of the FIM in the previous step). Such parameters are inherently unidentifiable, and should be fixed in subsequent steps to avoid computational cost. As in Step 1, we flag parameters as insensitive if their single parameter $\%CV \geq 100$. 
\\
\item \textbf{Parameter Combination Sets} Any remaining unidentifiable parameters  must be involved in identifiable combinations. In this step we determine which subsets of parameters form combinations in connected components. In general, we seek \emph{nearly full rank subsets}, i.e. rank-deficient subsets with the following property: for each parameter $p_i \in \mathcal{A}$ the rank of $\mathcal{A} \setminus p_i$ is full.  As a result, $\mathcal{A}$ will have exactly enough degrees of freedom for successful profile calculations. We also note that subsets satisfying this condition will always be within a single component, so that this partitions the parameter graph into connected components. This is because any subset $\mathcal{A}$ containing parameters from disconnected components $\mathcal{B}_1\dots\mathcal{B}_m$ must be full rank except for in a single component $\mathcal{B}_i$, to maintain an overall rank of $N-1$. But in this case, $\mathcal{A}$ will fail the property above as $\mathcal{A}\setminus p_j$ will not be full rank for $p_j \not \in \mathcal{B}_i$.
\\

These subsets can be determined in a number of ways, which are illustrated in Examples 2 and 3 in more detail. For example, one approach is to examine the rank of the FIM for increasing subsets of parameters (i.e. starting with pairs, then triplets, etc.). Any rank minimal rank deficient subsets must belong to the same connected component, so one can identify the connected components in this way and determine the rank of each component (number of identifiable combinations). We can then fix sufficiently many parameters in each component to make nearly full rank sets. Alternatively, one may also simply search for these nearly full rank sets directly, e.g. by testing the rank of decreasing sized parameter subsets beginning with $\mathcal{A}$. 
\\

\item \textbf{Likelihood Profiles} - Next we compute the profiles for the parameters in each subset, as these will force the profiles to trace out the form of the identifiable combinations. 
We consider only the flat regions of a the likelihood profile for each parameter $p_i$, and examine the relationship between $p_i$ and the other parameters. 
In general a threshold requiring the residual sum of squares to be less than $10^{-6}$ appears sufficient for most unidentifiable examples to make sure the profiles are taken in a flat region of likelihood space, although in practice the profiles in this case are often significantly smaller, e.g. on the order of $10^{-10}$.
\\


\item \textbf{Functional Form for the Identifiable Combinations} By fitting rational functions to the profile relationships, we determine an algebraic form for the relationship between $p^*$ and each fitted parameter. These curves represent the structurally identifiable combinations for the model, projected via evaluation maps for the remaining parameters to their fitted values (i.e. to values satisfying the identifiable combinations). 
From these relationships we can solve to recover the form for the overall identifiable combinations by combining these different projections together. While in practice for smaller models this is typically easy to do by inspection (see examples below), for more complicated models a more algorithmic approach or  multivariate polynomial interpolation approach may be preferable, which we aim to investigate further in future work. In general the combinations need not be rational functions, and in this case one could fit other combination functions to the profile relationships (e.g. exponential or piecewise functions); but we restrict to this case as it represents a commonly encountered class of models and there are existing methods for quickly interpolating rational functions from data. 
\end{enumerate}


\begin{figure}
\centering
\includegraphics[width=0.25\textwidth]{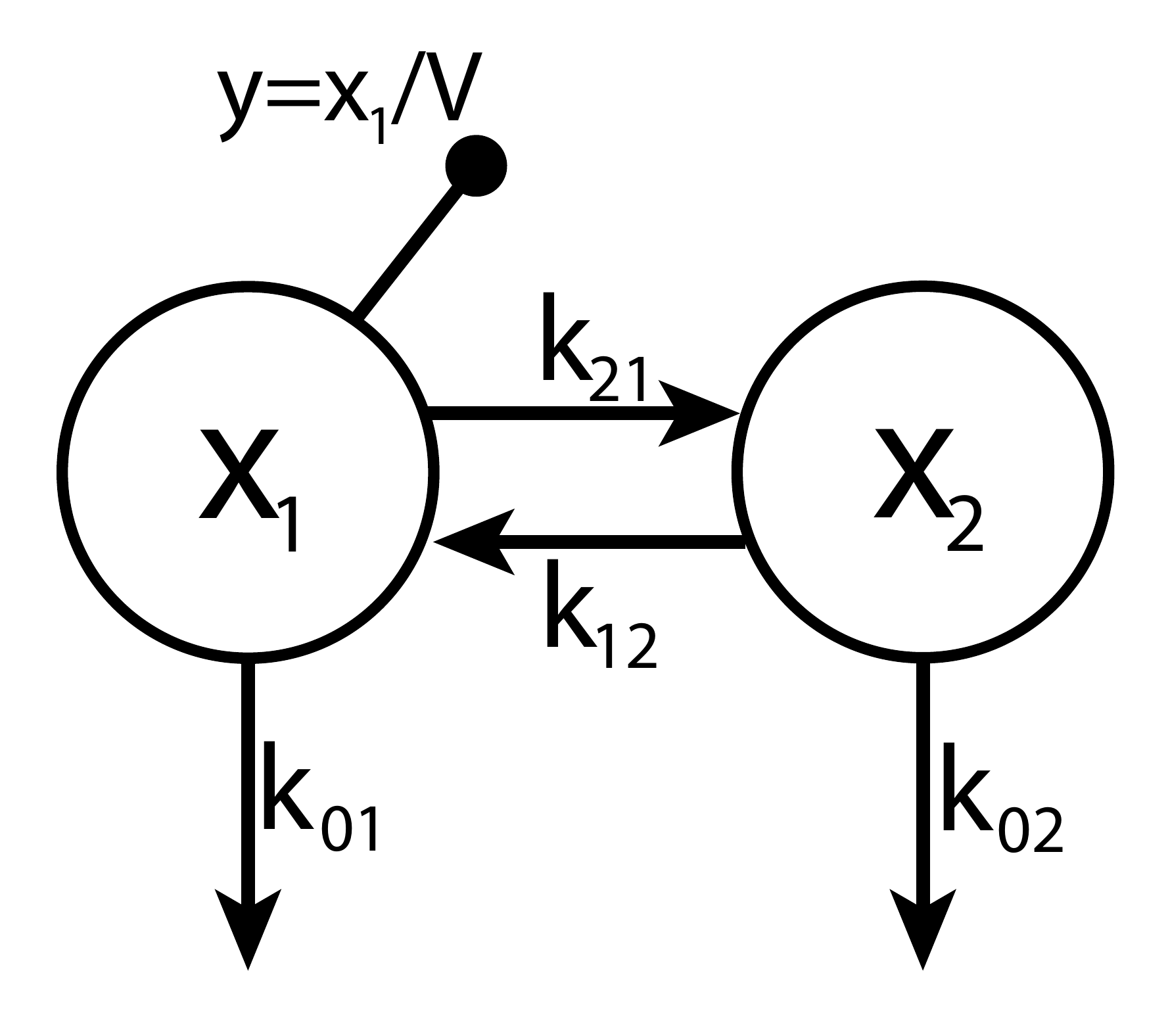} \hspace{2cm}
\includegraphics[width=0.28\textwidth]{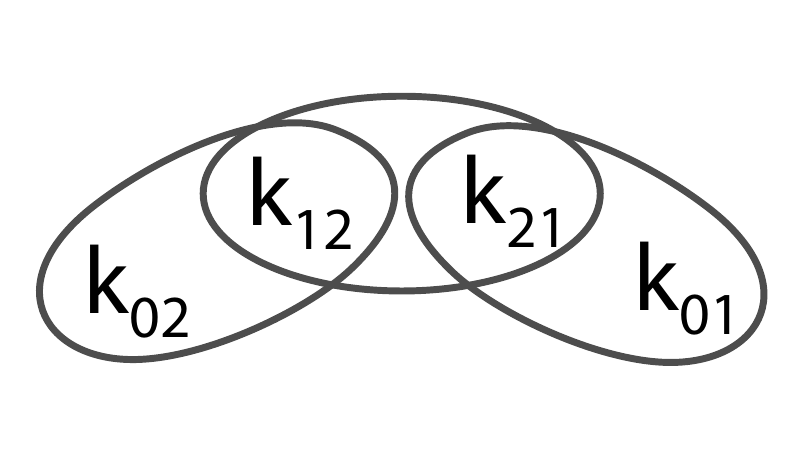}
\caption{Linear 2-compartment model diagram (left) and parameter graph (right). }
\label{fig:2comp}
\end{figure}

\section{Examples} \label{sec:exs}
In this section, we give several examples of the overall approach and illustrate some potential pitfalls. In Examples 1 - 3 we have chosen models where the identifiable combinations can also be determined analytically so as to compare our method with known results, and in Examples 4 and 5 we determine identifiable combinations for models with more complex nonlinearities. We implemented the method in Python 2.7, using Numpy and Scipy for numerical computation \cite{numpyscipy, scipy}. To fit rational functions in \textbf{Step 5}, we fit a series of increasing degree rational functions and used the Bayesian Information Criterion \cite{ward2008} to select the simplest among them. In practice, the resulting rational functions yielded near-perfect fits to the data, with sum of square residuals typically on the order of machine precision. 
\\

\begin{figure}
\centering
\includegraphics[width=0.23\textwidth]{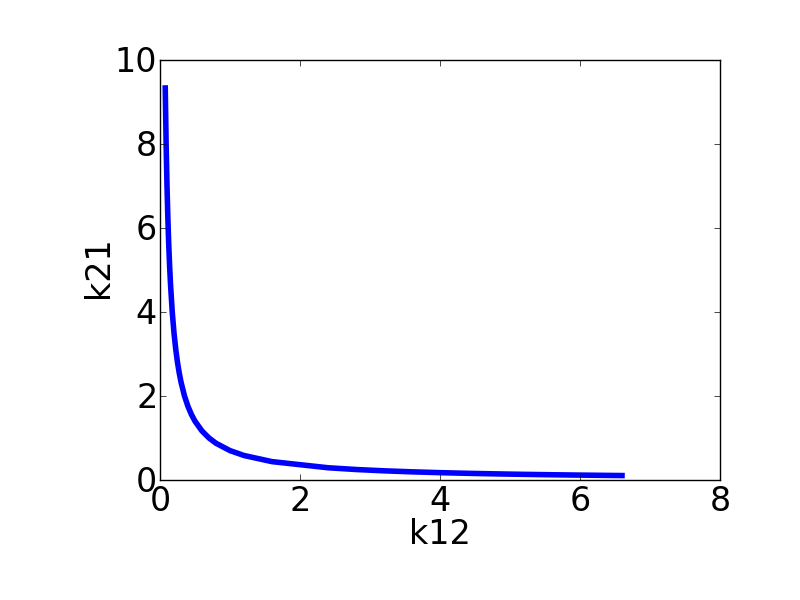}
\includegraphics[width=0.23\textwidth]{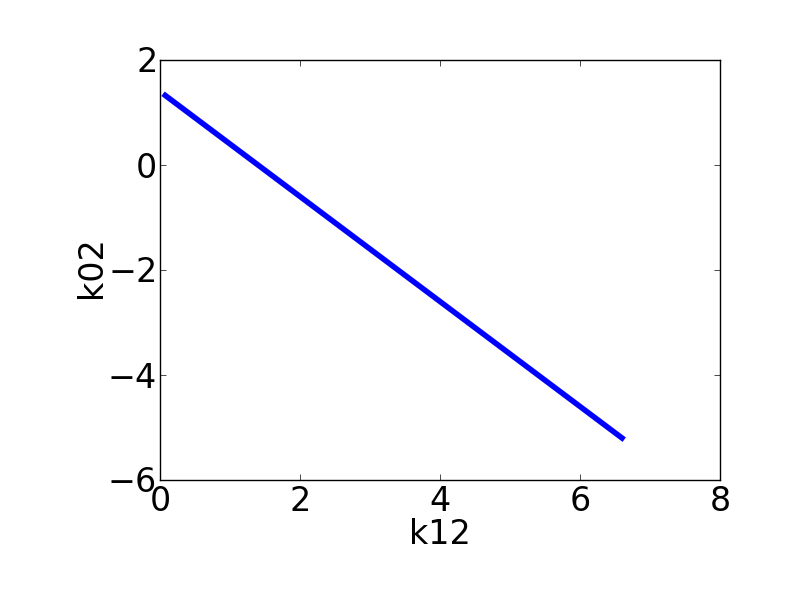}\\
\includegraphics[width=0.23\textwidth]{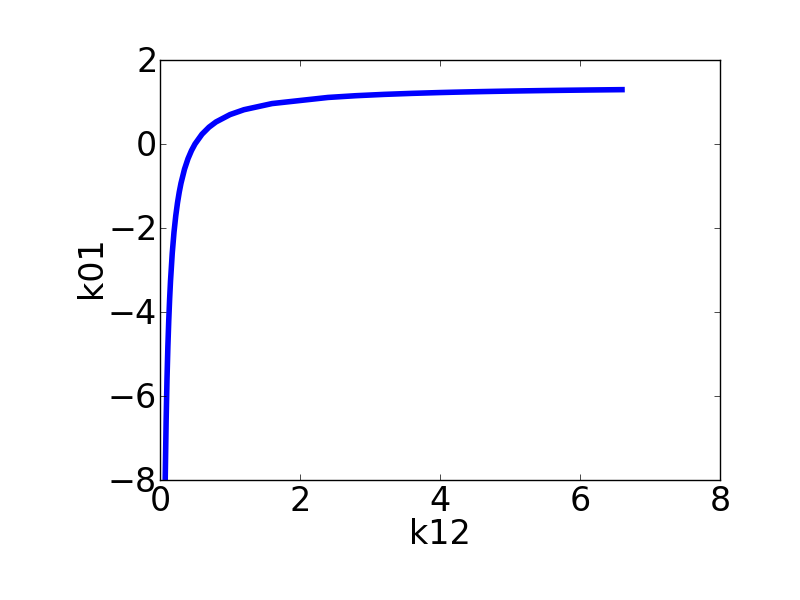}   
\includegraphics[width=0.23\textwidth]{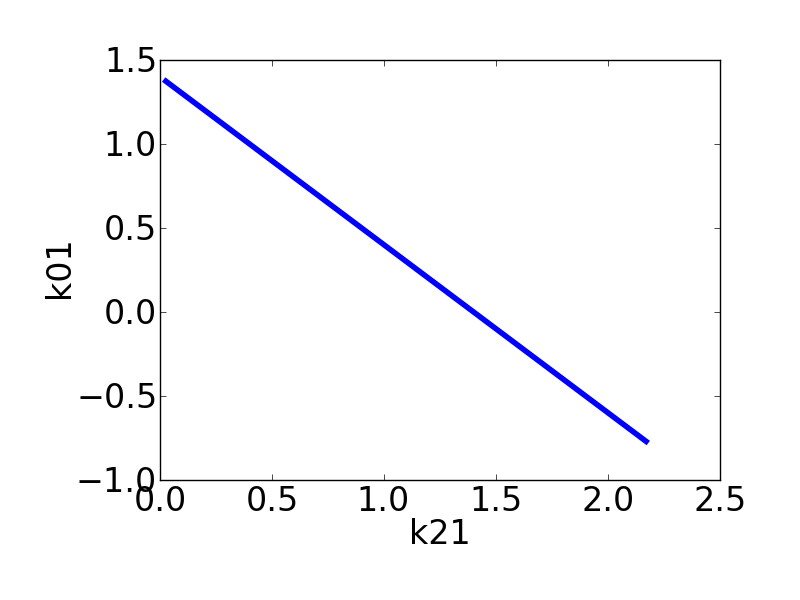}\\
\includegraphics[width=0.23\textwidth]{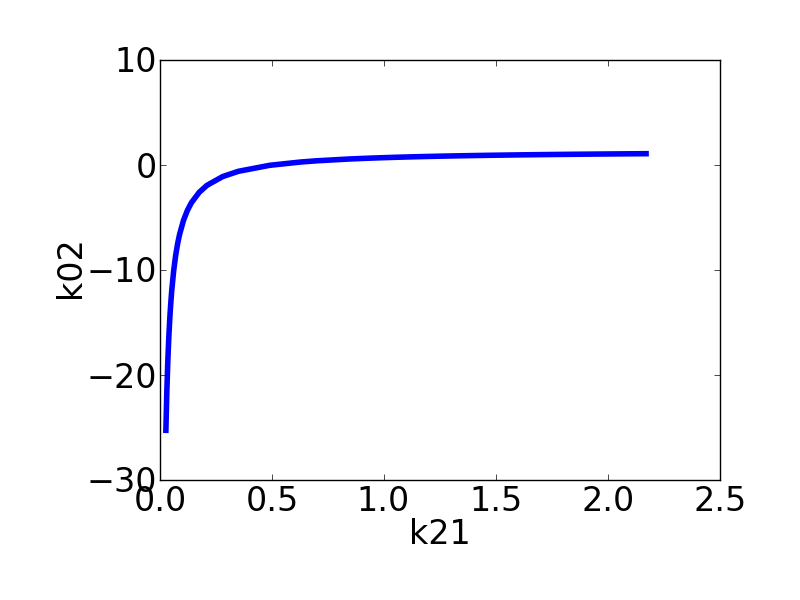}
\includegraphics[width=0.23\textwidth]{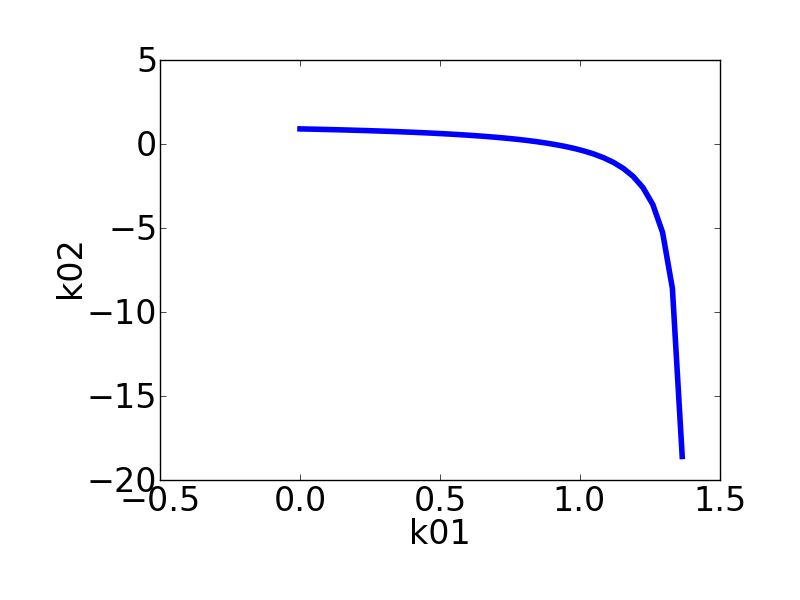}
\caption{Parameter relationships determined from likelihood profiles for the linear 2-compartment model in Example 1.}
\label{fig:2compprofiles}
\end{figure}

\noindent \textbf{Example 1: Linear 2-compartment Model}. The linear 2-compartment model (Figure \ref{fig:2compprofiles}) is commonly used in pharmacokinetic modeling, with equations given by:
\begin{equation}
\begin{aligned}
\dot{x}_1 &= k_{12} x_2 - (k_{01} + k_{21}) x_1 \\
\dot{x}_2 &= k_{21} x_1 - (k_{02} + k_{12}) x_2 \\
y &= x_1/V
\end{aligned}
\label{eq:2comp}
\end{equation}
where $x_1$ represents the mass of a substance in the blood (e.g. a hormone or drug), and $x_2$ represents the mass of the substance in the tissue.  The drug exchanges between blood and tissues, and is degraded/lost in both compartments, at the rates given by the $k_{ij}$'s above. 
The model output $y = x_1/V$ is the blood concentration of the drug, where $V$ is the blood volume. The $k_{ij}$'s and $V$ are unknown parameters to be estimated. 

This model has previously been shown to be unidentifiable using a range of analytical methods \cite{Meshkat2009, Audoly2001}, and the identifiable combinations are known to be $k_{12}k_{21}$, $k_{12}+k_{02}$, and $k_{21} + k_{01}$, with $V$ identifiable. As an example set of parameters, we take $k_{12} = 1, k_{21} = 0.7, k_{02} = 0.4, k_{01} = 0.7, V = 3$, and initial conditions $x_1(0) = 15, x_2(0) = 0$ (these parameters were chosen arbitrarily, however we tested a range of different parameters with similar results). 

Following the approach given above, in \textbf{Step 1}, we first note that the overall FIM is indeed rank deficient, with a rank of 4 (as there are 3 combinations and one identifiable parameter). The overall covariance matrix gives parameter \%CV's greater than $10^6\%$ for all $k_{ij}$'s, with the \%CV for $V$ of $4.78\%$, indicating that the individual $k_{ij}$'s are unidentifiable, but that $V$ is identifiable. In \textbf{Step 2}, we note that all individual parameter \%CV's (i.e. when fitting only a single parameter at a time) are all $<25\%$, indicating that no parameters are completely insensitive (i.e. they aren't inherently unidentifiable). Thus, all $k_{ij}$'s must be involved in identifiable combinations.  Indeed, the full set of $k_{ij}$'s satisfies the \textbf{Step 3} condition, indicating that the parameter combinations form a single connected component (Figure \ref{fig:2comp}). 

In \textbf{Step 4}, we generate profiles for each of the unidentifiable $k_{ij}$'s, and consider the relationship between the profiled and fitted parameters in the flat regions of the profile likelihood, shown in Figure \ref{fig:2compprofiles}. The likelihood profiles were consistently flat over the parameter region tested with a residual sum of squares $<10^{-10}$ for all parameters. In \textbf{Step 5}, fitting to the curves in Figure \ref{fig:2compprofiles} gives the following relationships (after clearing denominators and factoring): 
\begin{equation}
\begin{aligned}
k_{12} k_{21} &= 0.7\\
k_{12} + k_{02} &= 1.4\\
k_{12}(1.4 - k_{01}) &= 0.7\\
k_{21}+k_{01} &= 1.4\\
k_{21}(1.4 - k_{02}) &= 0.7\\
(1.4 - k_{01})(1.4 - k_{02}) &= 0.7
\end{aligned}
\label{eq:2compprofile}
\end{equation} 
From the rank of the original FIM, we expect to find 3 identifiable combinations (as the rank was 4 and $V$ is identifiable), and indeed by inspection, we can see that the identifiable combinations consistent with Eq. \eqref{eq:2compprofile} are $k_{12}k_{21}$, $k_{12}+k_{02}$, and $k_{21} + k_{01}$. 
\\
\\

\noindent \textbf{Example 2: 2-compartment model with a rank-deficient parameter pair}. To illustrate how \textbf{Step 3} works when the number of parameters is greater than the number of combinations plus one, we consider the following simple variant of the previous example:
\begin{equation}
\begin{aligned}
\dot{x}_1 &= k_{1} x_2 - (k_{2} + k_{3} + k_{4}) x_1 \\
\dot{x}_2 &= k_{4} x_1 - (k_{5} + k_{1}) x_2 \\
y &= x_1/V
\end{aligned}
\label{eq:loose}
\end{equation}
which is equivalent to Eq. \eqref{eq:2comp} with $k_{12} = k_1, k_{01} = k_2 + k_3, k_{21} = k_4$, and $k_{02} = k_5$. The identifiable parameter combinations are thus $k_1k_4, k_2 + k_3+k_4$, and $k_1+k_5$, with $V$ again identifiable. A diagram of these combinations is given as the bottom component of the example in Figure \ref{fig:looseparams}.  Based on these combinations, note that one of $k_2$ and $k_3$ must be fixed when profiling in order to yield an identifiable model (as there are 3 combinations but 5 $k$'s). As an example set of parameters, we take $k_1 = 2.3, k_2 = 0.421, k_3 = 0.52, k_4 = 0.61, k_5 = 1.23$, and $V = 2.2$ (as before, the results are similar for a range of parameter values). 

\begin{figure}
\centering
\includegraphics[width=0.35\textwidth]{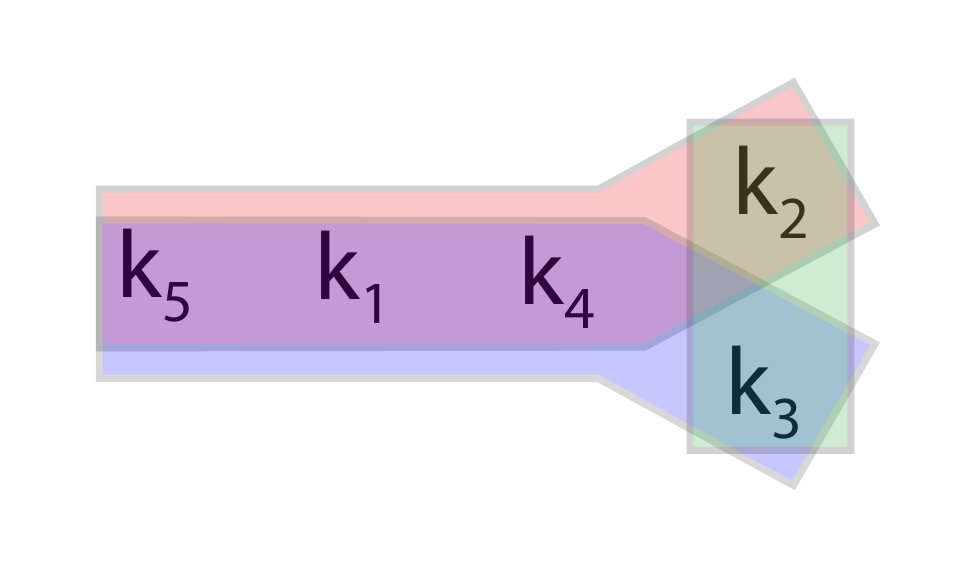}
\caption{Parameter combinations for the model in Example 2. Nearly full rank subsets shaded, as determined by subset rank search. }
\label{fig:looserankparams}
\end{figure}

The results for \textbf{Steps 1} and \textbf{2} are similar to Example 1.  The full model FIM has rank 4 and $V$ is identifiable so we expect 3 identifiable combinations.  When we consider the subsets of parameters in \textbf{Step 3}, we find that  $\left\{k_5, k_1, k_4, k_2 \right\}$, $\left\{k_5, k_1, k_4, k_3 \right\}$, and $\left\{ k_2, k_3 \right\}$ satisfy our criteria.  These subsets form a single connected component with a loose pair (Figure \ref{fig:looserankparams}).  Note that the loose pair, $\left\{ k_2, k_3 \right\}$ is the only rank deficient pair.  As a result, any subset including both of these parameters cannot satisfy the condition in \textbf{Step 3}.

\begin{figure}
\centering
\includegraphics[width=0.4\textwidth]{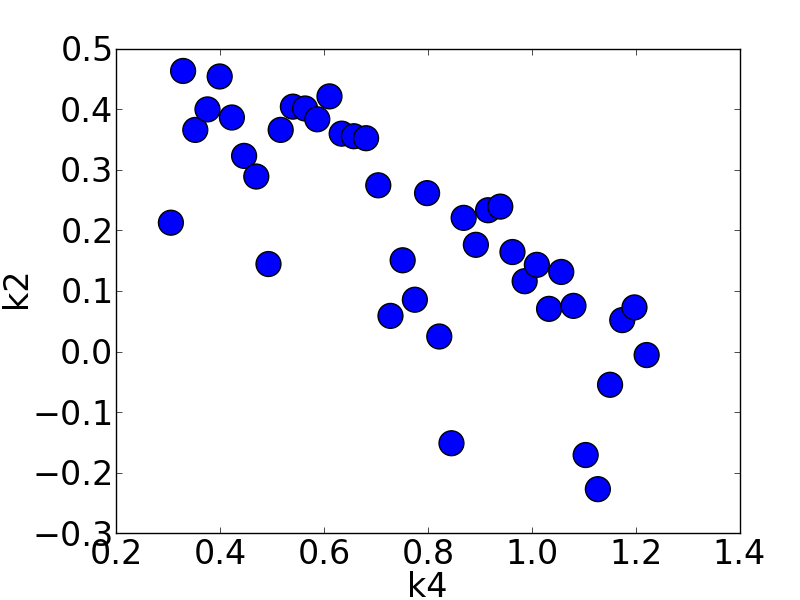}
\caption{Example parameter relationship for $k_4$ and $k_2$ when there are loose parameters in the profile fit (i.e. there are more parameters than degrees of freedom) The resulting parameter relationship shows no precise relationship between $k_4$ and $k_2$, even though the likelihood is flat in this region. We note that the general trends of the relationships between parameters can still be seen because we're starting close to the true values in each step of the profile, but the specific form of the identifiable combinations can't be determined.}
\label{fig:notfixed}
\end{figure}

To illustrate the necessity of \textbf{Step 3}, Figure \ref{fig:notfixed} shows an example result of the relationship between $k_4$ and $k_2$ if all parameters except $k_4$ are fitted. As $k_4$ is shifted along the $x$-axis, $k_2$ and $k_3$ are not fully constrained, i.e. they both may take on any values that maintain $k_2+k_3 = 0.941-k_4$. This results in the appearance of a scatterplot in Figure \ref{fig:notfixed} with no clear relationship between $k_4$ and $k_2$, in spite of the fact that they are part of an identifiable combination.

\begin{figure*}
\centering
\includegraphics[width=0.32\textwidth]{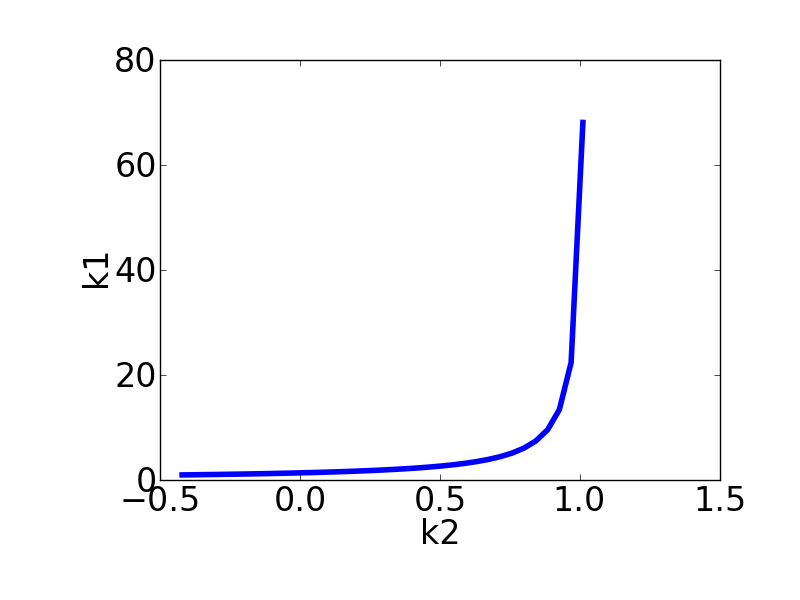}
\includegraphics[width= 0.32\textwidth]{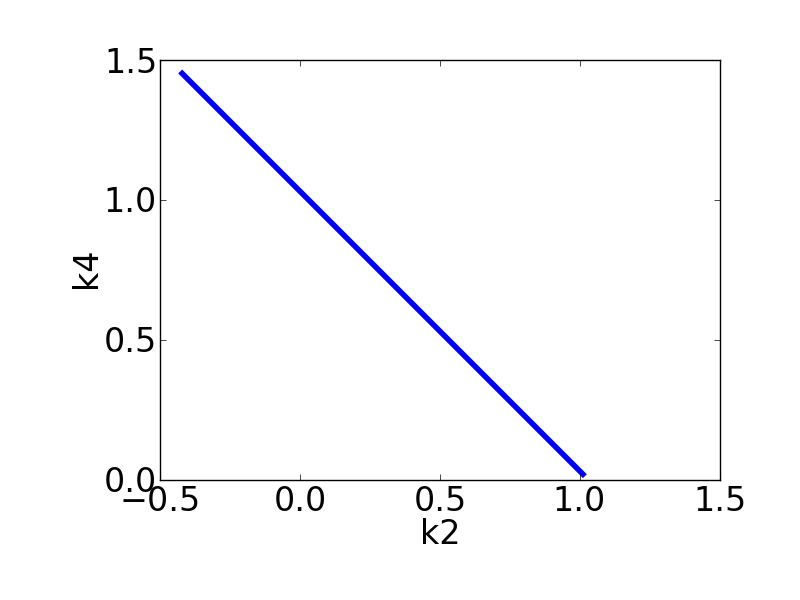}
\includegraphics[width= 0.32\textwidth]{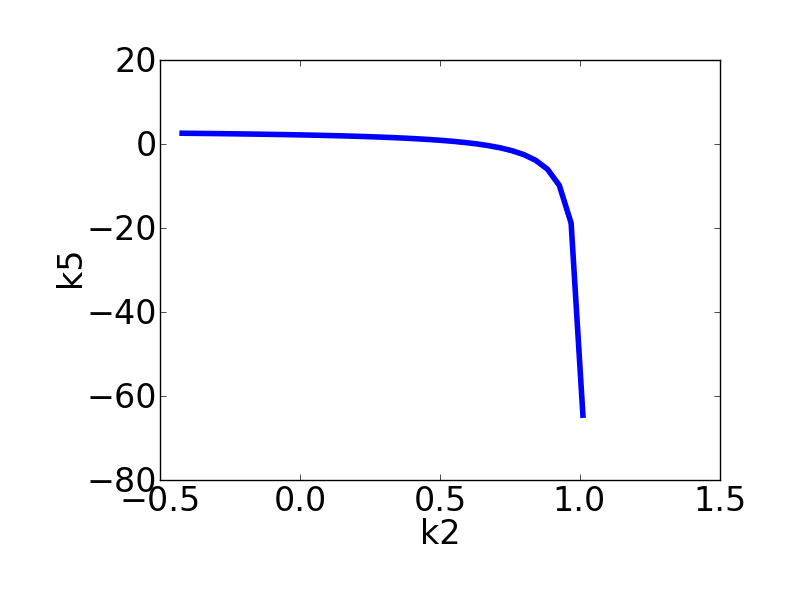}\\
\includegraphics[width= 0.32\textwidth]{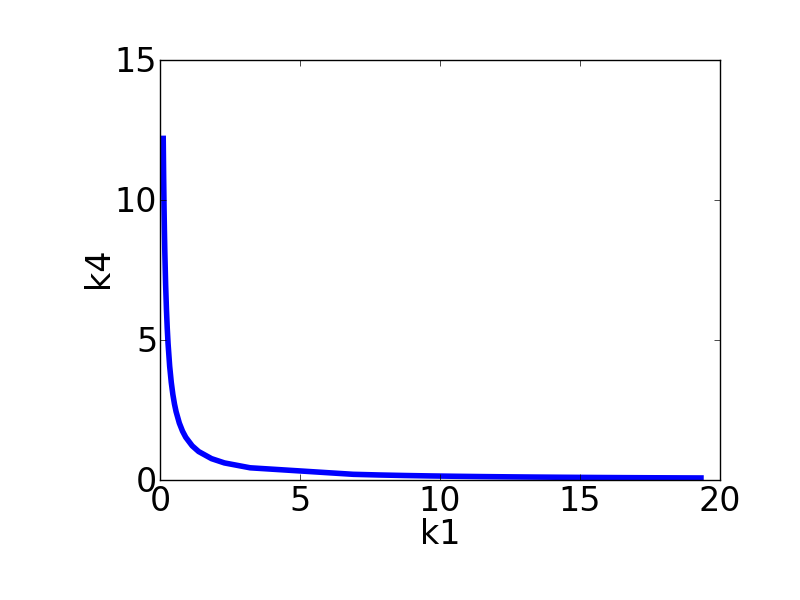}
\includegraphics[width= 0.32\textwidth]{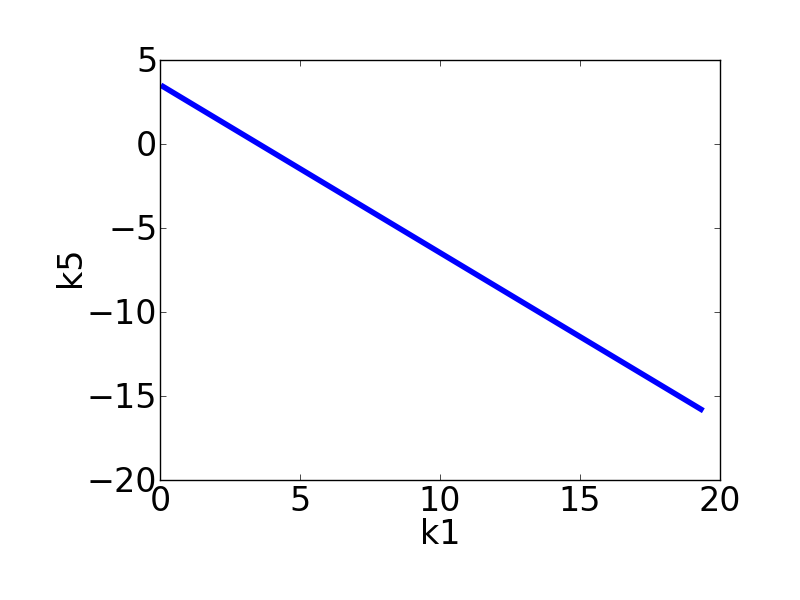}
\includegraphics[width= 0.32\textwidth]{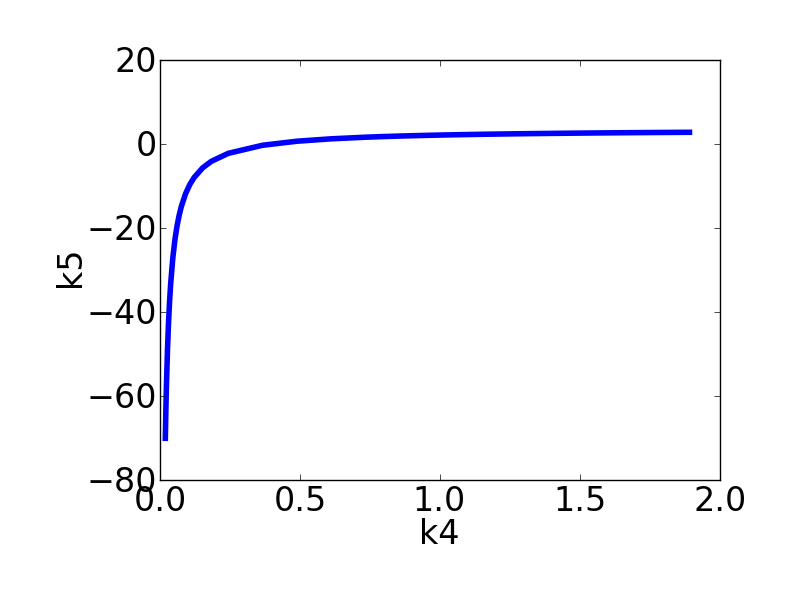}\\
\includegraphics[width= 0.32\textwidth]{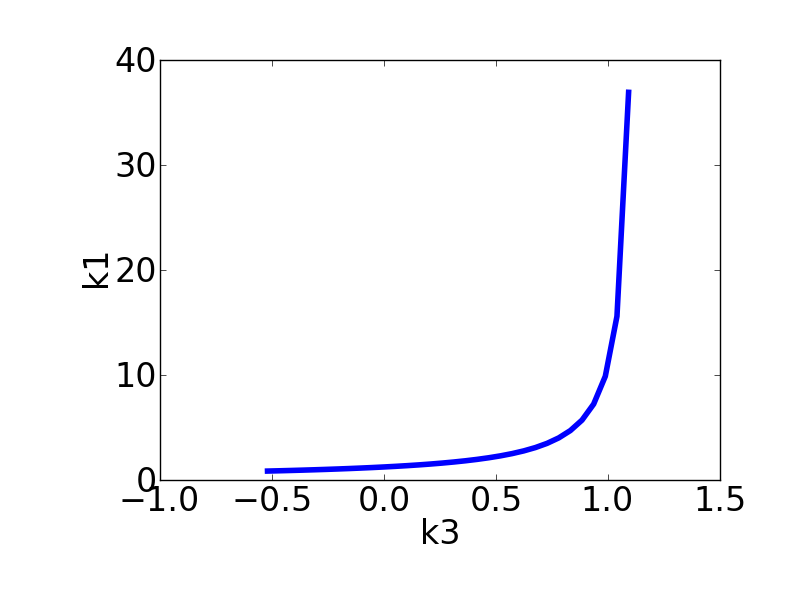}
\includegraphics[width= 0.32\textwidth]{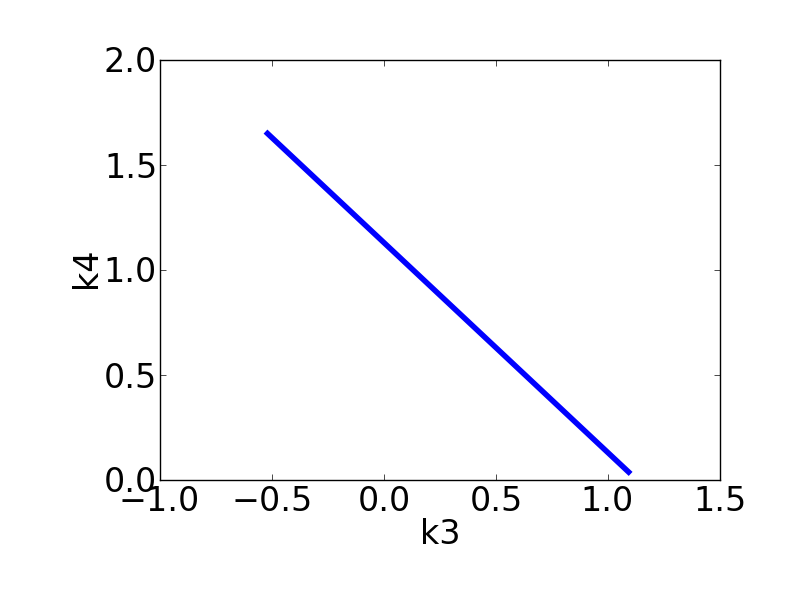}
\includegraphics[width= 0.32\textwidth]{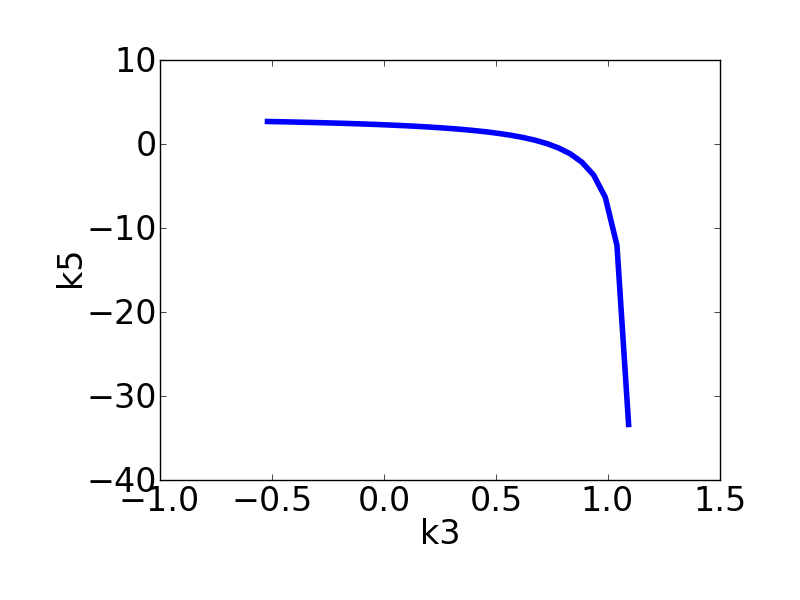}\\
\includegraphics[width= 0.32\textwidth]{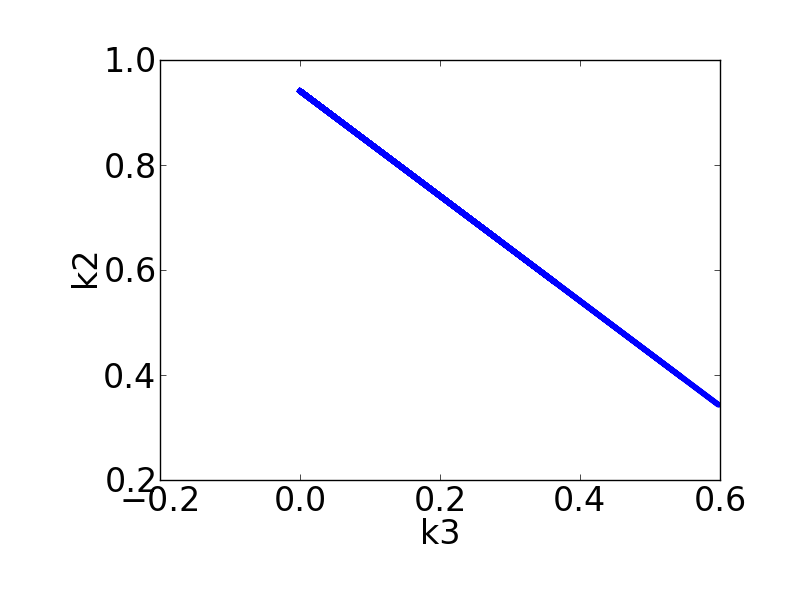}
\caption{Parameter relationships determined from likelihood profiles for Example 2.}
\label{fig:looseprofiles}
\end{figure*}

Thus, in \textbf{Step 4}, we profile parameters within each subset, fitting only the remaining parameters in the subset.  This results in 10 distinct parameter relationships from the likelihood profiles, shown in Figure \ref{fig:looseprofiles}.  For efficiency, it is not necessary to profile parameters twice where subsets overlap to capture all pairwise relationships.  That is, assuming parameters in subsets $\left\{k_5, k_1, k_4, k_2 \right\}$ and $\left\{k_2, k_3\right\}$ have been profiled, we need only compute a profile for $k_3$ in the remaining subset.  

In \textbf{Step 5}, rational function fitting of the parameter relationships in Figure \ref{fig:looseprofiles} yields the following equations:
\begin{equation}
\begin{aligned}
k_1 &= \frac{1.403}{1.031 - k_2}\\
k_4 &= 1.031 - k_2\\
k_5 &= \frac{2.23643 - 3.53 k_2}{1.031 - k_2}\\
k_4 &=\frac{1.403}{k_1}\\
k_5 &=3.53 - k_1\\
k_5 &= 3.53 - \frac{1.403}{k_4}\\
k_1 &= \frac{1.403}{1.13 - k_3}\\
k_2 &= 0.941 - k_3\\
k_4 &=1.13 - k_3\\
k_5 &= \frac{2.5859 - 3.53 k_3}{1.13 - k_3}
\end{aligned}
\label{eq:looseprofile}
\end{equation} 
From the second, fourth, fifth, eighth, and ninth equations above, we see that $k_2 + k_3, k_3+k_4, k_2 + k_4, k_1+k_5$, and $k_1k_4$ must be terms within our identifiable combinations. As we expect to have 3 identifiable combinations, we can see from these expressions that our identifiable combinations are most likely $k_2+k_3+k_4, k_1+k_5$, and $k_1k_4$. Testing this against the remaining equations in Eq. \eqref{eq:looseprofile} shows that indeed these are the identifiable combinations, which matches the combinations found analytically above. 
\\

\begin{figure*}
\centering
\includegraphics[width=0.29\textwidth]{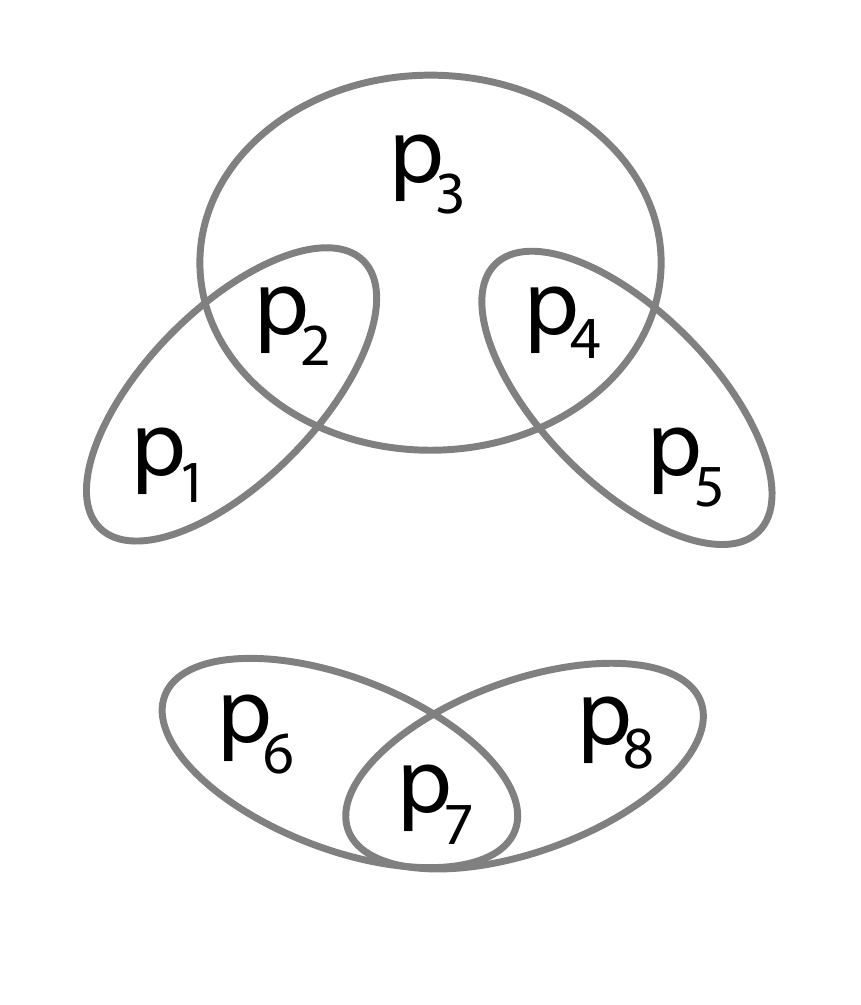} \hspace{0.05\textwidth}
\includegraphics[width=0.29\textwidth]{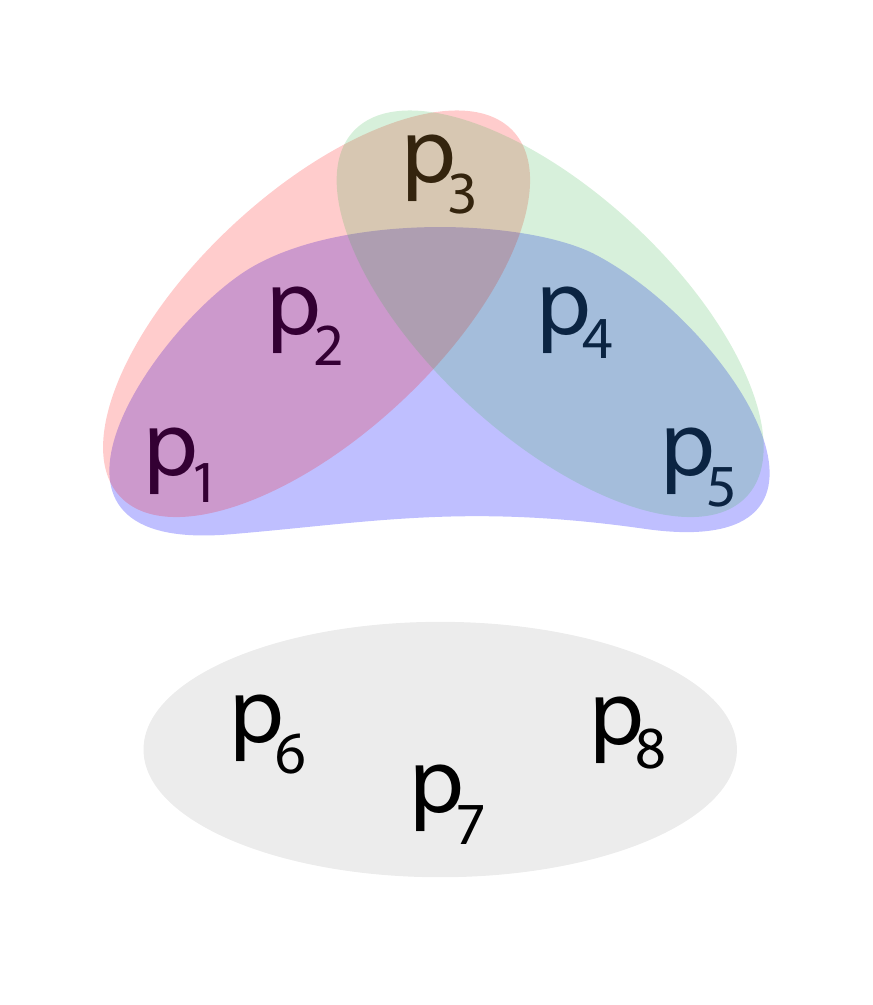}
\caption{Parameter combinations for the model in Example 3. Left: Grey circles indicate identifiable parameter combinations, e.g. $p_1$ and $p_2$ are involved in an identifiable combination. Some parameters are involved in more than one identifiable combination (e.g. $p_2$), so that there are two overall connected components. Right: Nearly full rank subsets (shaded) and overall connected components as determined by subset rank search.}
\label{fig:multicompparams}
\end{figure*}

\pagebreak

\noindent \textbf{Example 3: Multiple connected components of parameter combinations}. Next, we demonstrate the method using a simple example that gives multiple connected components of parameters. In this case, we will use a nonlinear variation of the two compartment model, which has been shown previously to be structurally identifiable \cite{Audoly2001}. The model equations are as follows:
\begin{equation}
\begin{aligned}
\dot{x}_1 &= k_{12} x_2 - \frac{V_{max} x_1}{K_m + x_1} - k_{21} x_1\\
\dot{x}_2 &=  k_{21} x_1 - (k_{12} + k_{02}) x_2\\
y &= x_1
\end{aligned}
\label{eq:multicomp}
\end{equation}
To generate an unidentifiable model with parameter combinations which form multiple connected components, we take $k_{12} = p_1 p_2$, $V_{max} = p_2+p_3+p_4$, $K_m = p_4+p_5$, $k_{21} = p_6+p_7$, and $k_{02} = p_7 + p_8$, and let $p_1, \dots, p_8$ be our parameters to be estimated. As the original model is identifiable, these forms are also our identifiable combinations. Figure \ref{fig:multicompparams} shows a diagram of the connected parameter components of this model. As an example set of parameter values, we take $p_1 = 2, p_2 = 0.4, p_3 = 3, p_4 = 0.8, p_5 = 1.2, p_6 = 0.8, p_7 = 1.5$, and $p_8 = 0.3$. 

In \textbf{Step 1}, the full FIM gives rank 5, correctly indicating that we expect to have 5 identifiable combinations, and gives each of the parameter \%CV's on the order of $10^5$-$10^8$, indicating that all the individual parameters are unidentifiable. The single-parameter \%CV's in \textbf{Step 2} are all $<10\%$, so we know that none of the parameters are completely insensitive, and thus are all likely to be involved in identifiable combinations. 

In \textbf{Step 3}, we search the subsets in order of decreasing size to find nearly full rank subsets for estimation. Namely, the subsets $\left\{p_1, p_2, p_4, p_5\right\}$, $\left\{p_1, p_2, p_3\right\}$, $\left\{p_3, p_4, p_5\right\}$ and $\left\{p_6, p_7,p_8\right\}$ satisfy our selection criteria (Figure \ref{fig:multicompparams}).   The first three subsets share parameters, indicating that $\left\{p_1, p_2, p_3, p_4, p_5 \right\}$ form a connected component while $\left\{p_6, p_7, p_8 \right\}$ form another.  

\begin{figure*}
\centering
\includegraphics[width=0.32\textwidth]{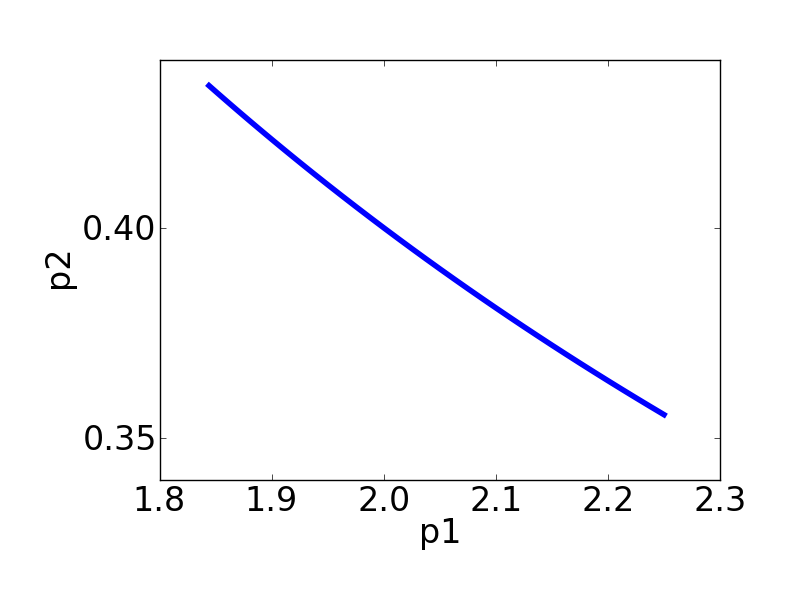}
\includegraphics[width=0.32\textwidth]{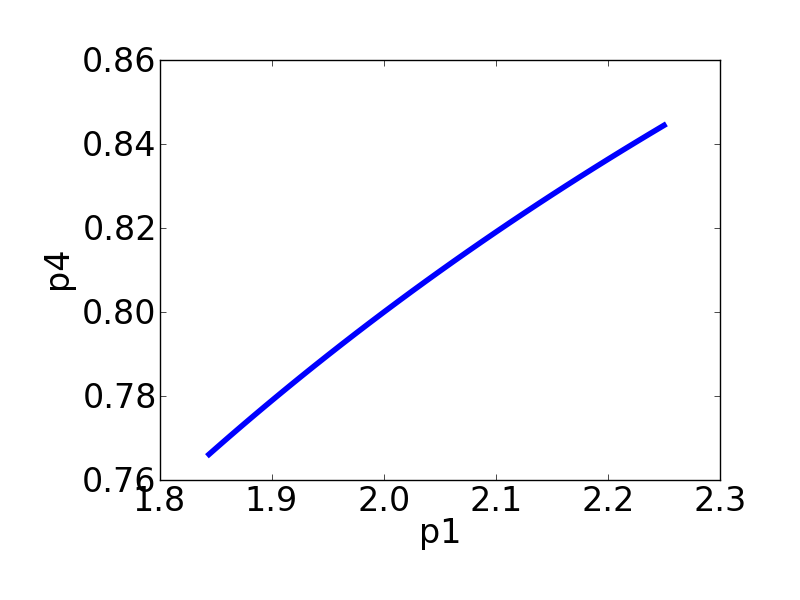}
\includegraphics[width=0.32\textwidth]{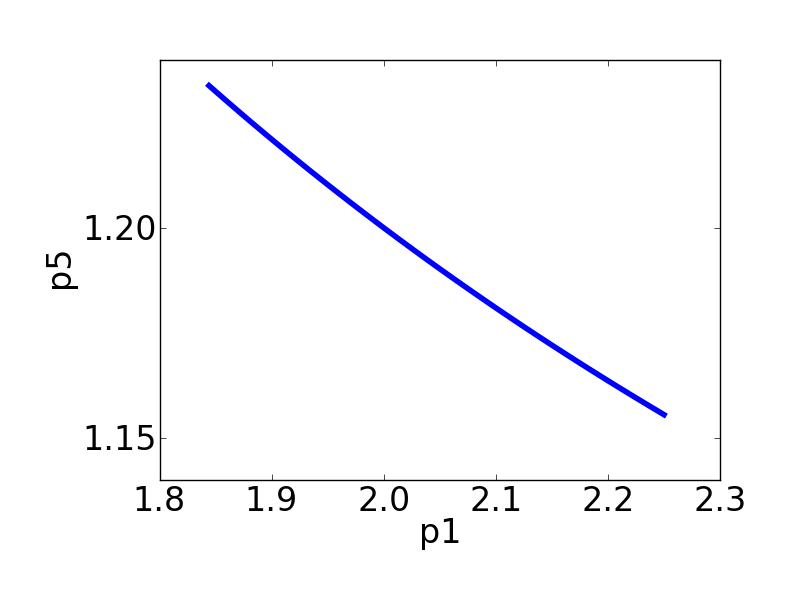}\\
\includegraphics[width=0.32\textwidth]{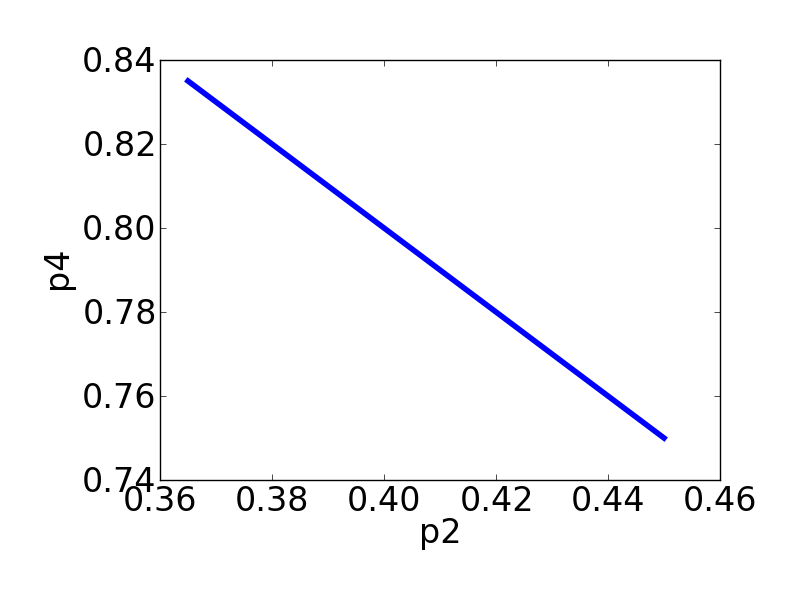}
\includegraphics[width=0.32\textwidth]{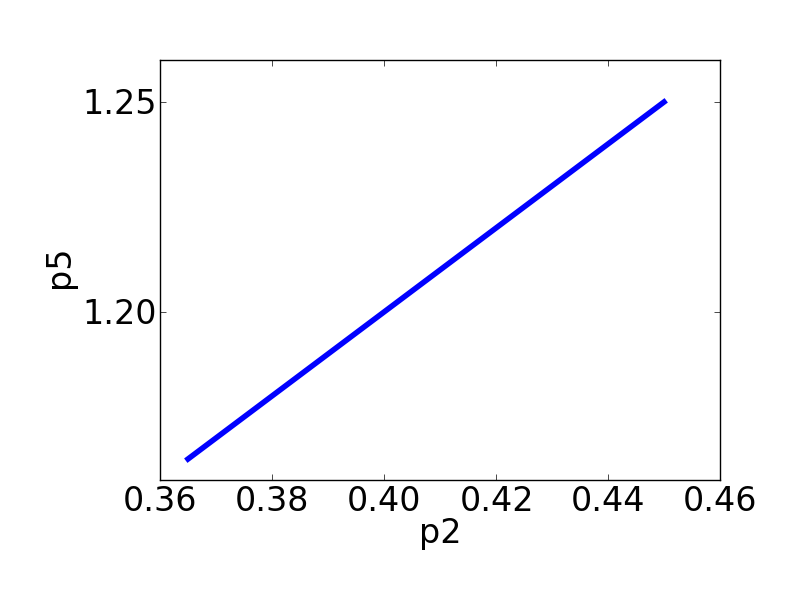}
\includegraphics[width=0.32\textwidth]{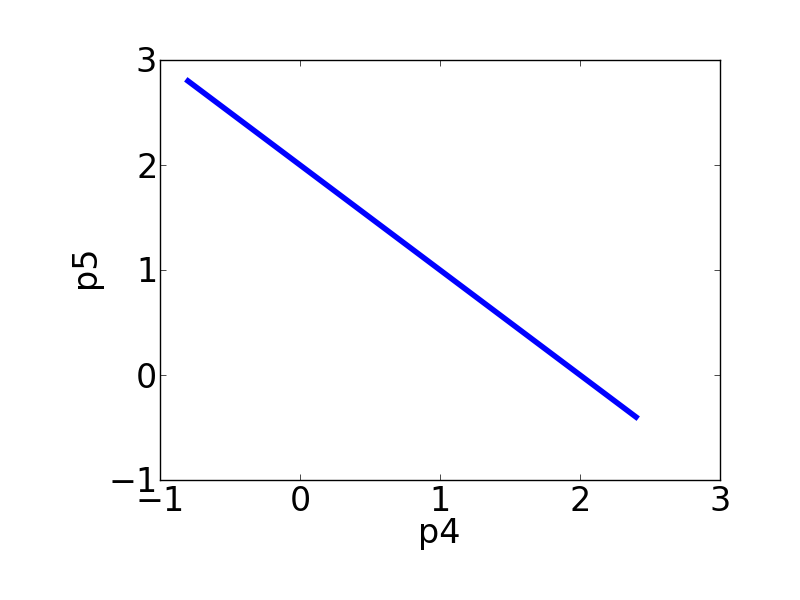}\\
\includegraphics[width=0.32\textwidth]{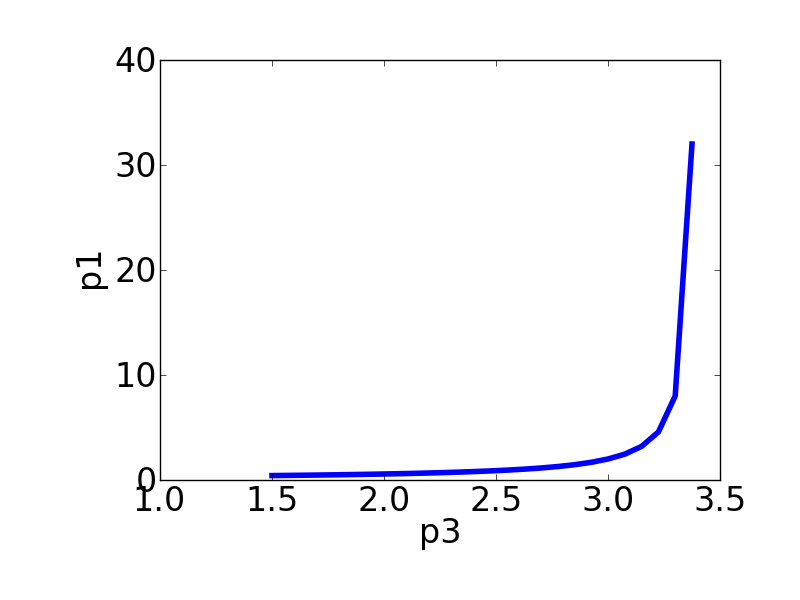}
\includegraphics[width=0.32\textwidth]{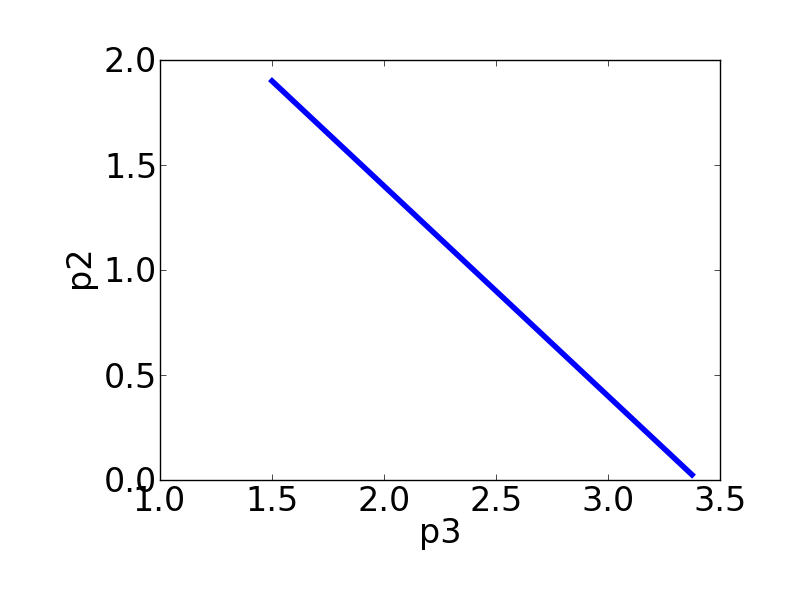}
\includegraphics[width=0.32\textwidth]{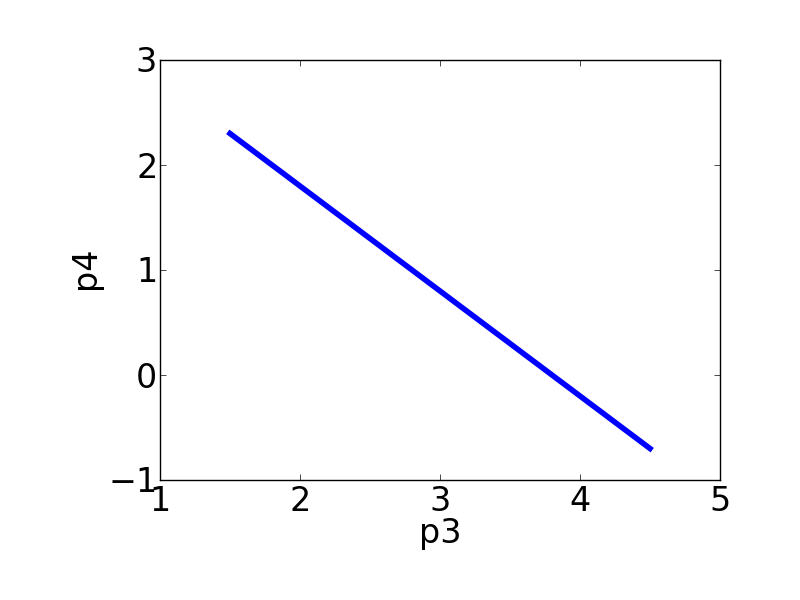}\\
\includegraphics[width=0.32\textwidth]{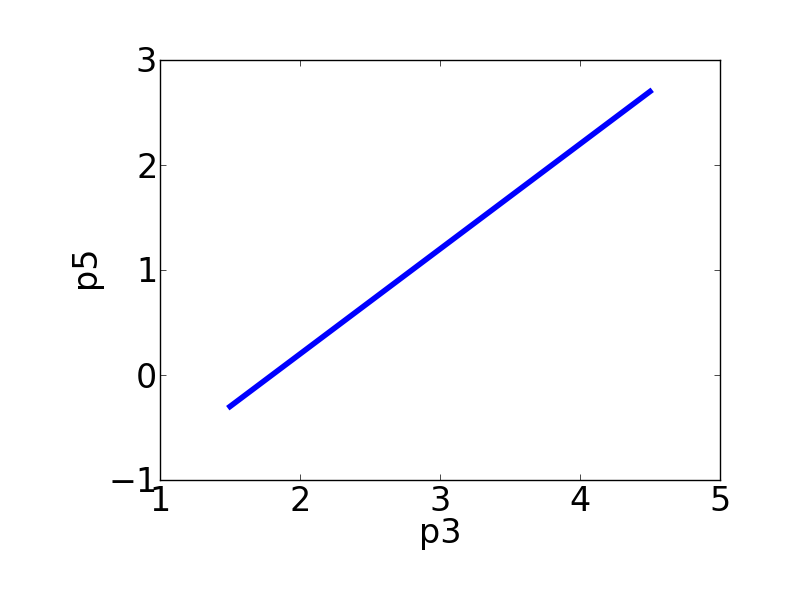}
\includegraphics[width=0.32\textwidth]{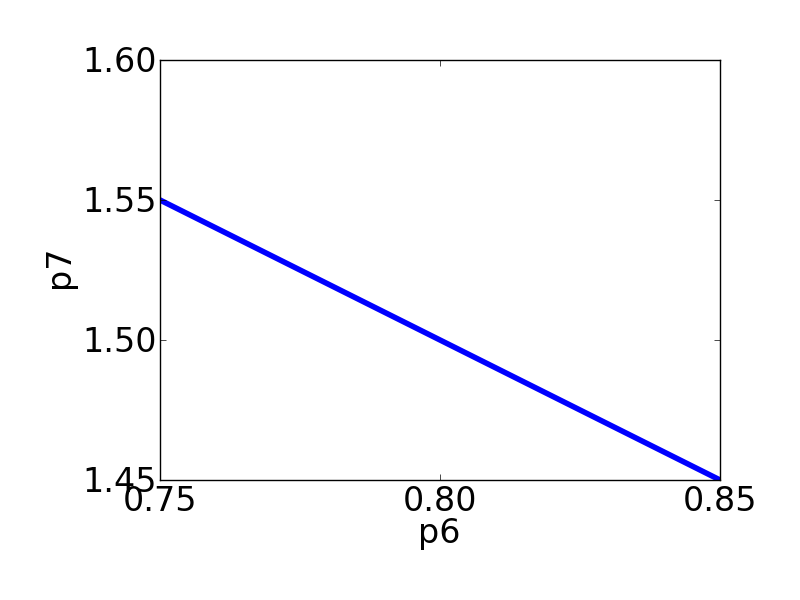}
\includegraphics[width=0.32\textwidth]{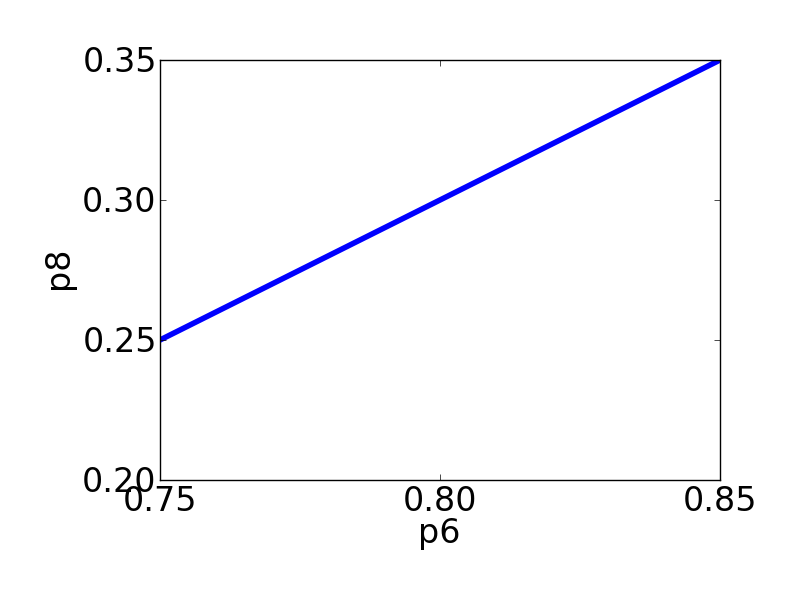}\\
\includegraphics[width=0.32\textwidth]{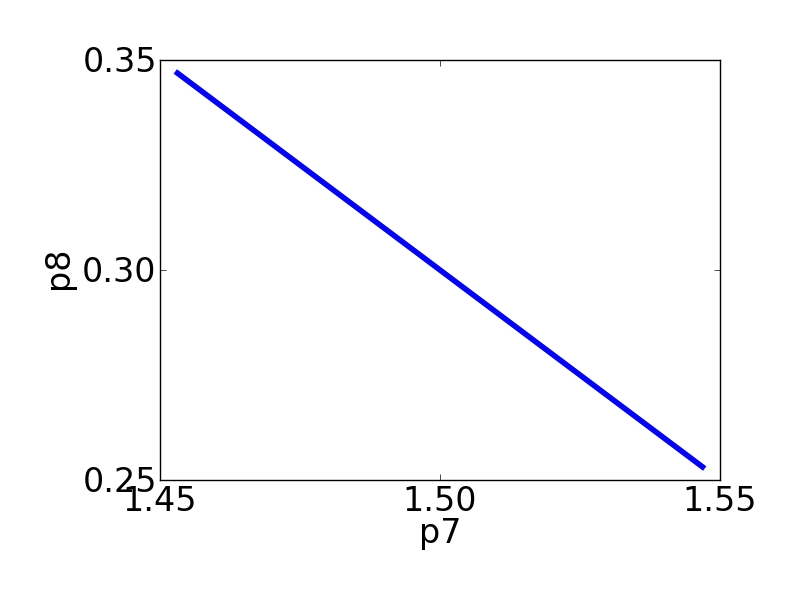}
\caption{Parameter relationships determined from likelihood profiles for Example 3.}
\label{fig:multicompprofiles}
\end{figure*}


We use these subsets for the profile likelihoods in \textbf{Step 4}, noting that the subsets capture all 13 unique pairwise parameter relationships (Figure \ref{fig:multicompprofiles}). Rational function fitting in \textbf{Step 5} yields the following parameter relationships: 
\begin{equation}
\begin{aligned}
p_2 &= \frac{0.8}{p_1}\\
p_4 &= 1.2 - \frac{0.8}{p_1}\\
p_5 &= -0.8+\frac{0.8}{p_1}\\
p_4 &= 1.2 - p_2\\
p_5 &= p_2 + 0.8\\
p_5 &= 2-p_4\\
p_1 &= \frac{0.8}{3.4-p_3}\\
p_2 &= 3.4 - p_3\\
p_4 &=3.8 - p_3\\
p_5 &=-1.8 + p_3\\
p_7 &= 2.3 - p_6\\
p_8 &=-0.5 +p_6\\
p_8 &=1.8 - p_7
\end{aligned}
\label{eq:multicompprofile}
\end{equation} 
For the first component, we examine the first ten equations above. From the first, fourth, fifth, sixth, eighth, ninth, and tenth equations, we see that we would expect $p_1 p_2$, $p_4+p_2$, $p_5-p_2$, $p_5+p_4$, $p_2+p_3$, $p_4+p_3$, and $p_5-p_3$ to be parts of various combinations. As this component has rank 3, we expect these to collapse to form 3 combinations. We first propose $p_1 p_2$ as an identifiable combination. For the remaining paired sums and differences, there are several equivalent ways we can collapse them into combinations. For example, as all three pairs $p_2+p_3$, $p_4+p_3$, $p_4+p_2$ appear in the list of functions, we propose the sum $p_2 + p_3 +p_4$ as a combination. This then leaves $p_5-p_2$, $p_5+p_4$, and $p_5-p_3$, all of which can be explained by letting $p_5+p_4$ be a combination. This set of three combinations $p_1p_2$, $p_2 + p_3 +p_4$, and $p_5+p_4$ is consistent with all the profiled parameter relationships in this component, gives the appropriate rank of 3, and matches the analytically determined identifiable combinations above. Alternatively, one could have collapsed the pairwise sums into combinations such as $p_5 - p_2 - p_3$ and $p_4 + p_5$, which would also give the appropriate rank and match the profiled relationships in \eqref{eq:multicompprofile}.

From the last three equations, we can see that we expect $p_6+p_7$, $p_8 - p_6$, and $p_8+p_7$ to be included as pieces in the combinations. This component has rank 2, and indeed we see that there are only two algebraically independent combinations among these three (e.g. $p_8 - p_6$ is the difference of the other two expressions), so that we can take our combinations to be $p_6+p_7$ and $p_8 + p_7$. 
%
\\
\\

\begin{table}\centering \small 
\begin{tabular}{ccl}
\textbf{Parameter} & \textbf{Value} & \textbf{Description}\\
\hline
$B_0$ & 1.166 & Basal TSH secretion \\
$A_0$ & $\frac{581}{1166}$ & TSH rhythm amplitude \\
$c$ & 1 & Damping coefficient \\
$k_{34}$ & 0.118 & Brain influx and conversion rates for $T_3$ and $T_4$ \\
$k_{degTSH}$ & 0.756 & TSH degradation rate \\
$k_{degT_{3B}}$ & 0.037 & $T_{3B}$ degradation rate 
\end{tabular}
\caption{Parameter values and descriptions for the TSH model in Example 4.}
\label{tab:TSH}
\end{table}

\noindent \textbf{Example 4: Modeling Thyroid Hormone Dynamics}.  The following example demonstrates the ability of our approach to examine models that include non-rational functions.  Eisenberg and DiStefano's thyroid hormone model \cite{Eisenberg2008} includes a sine and exponential function, which are outside the scope of the standard differential algebra approach.  The model equations are given below.
\begin{equation}
\begin{aligned}
\dot{TSH} &= 1000B_0(1 + A_0 sin(\frac{2\pi}{24}t - \phi))e^{-cT_{3B}}- k_{degTSH}TSH\\
\dot{T}_{3B} &= k_{34}(T_{3P} + T_{4P}) - k_{degT_{3B}}T_{3B}\\
y &= TSH
\end{aligned}
\label{eq:tsh}
\end{equation}
where $TSH$  is the mass of thyroid stimulating hormone, $T_{3B}$ and $T_{3P}$ are the masses of triiodothyronine in the brain and plasma, respectively, and $T_{4P}$ is the mass of thyroxine in the plasma.  We assume that $T_{3P}$ and $T_{4P}$ are measurable, so $T_{3P} + T_{4P}$ is a known quantity.  Table \ref{tab:TSH} defines the model parameters.  

\begin{figure}
\centering
\includegraphics[width=0.33\textwidth]{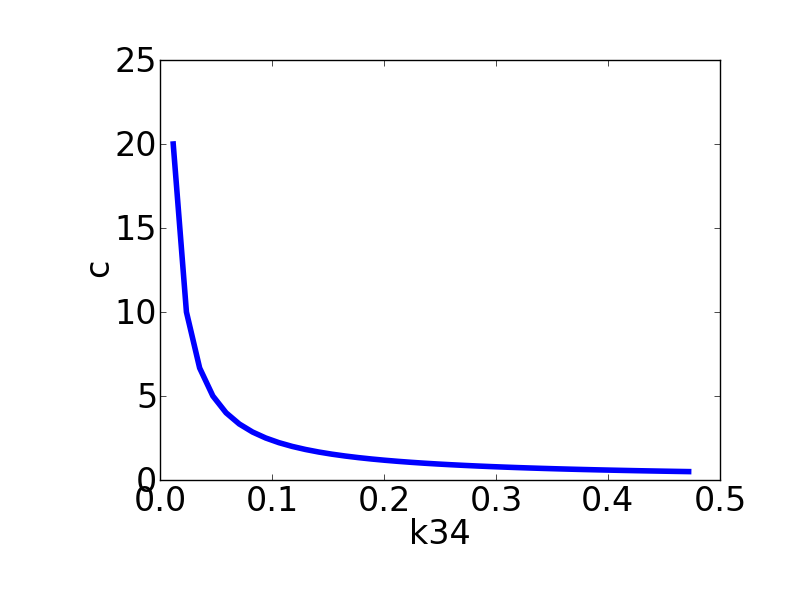}
\caption{Parameter relationship determined from likelihood profile for Example 4.}
\label{fig:tshprofile}
\end{figure}

In \textbf{Step 1}, the FIM generated using all seven parameters has rank 6.  In addition, $\phi$, $k_{degTSH}$, $k_{degT3B}$, $A_0$, and $B_0$ have \%CV's $< 25$, indicating that these parameters are identifiable.  The single parameter \%CV's computed in \textbf{Step 2} are all on the order of $10^{-2}$, so no parameters are insensitive.  \textbf{Step 3} confirms that the remaining parameters $(k_{34}, c)$ form a rank deficient pair that is also the only connected component in the parameter graph.  For \textbf{Step 4} we profile $k_{34}$, fitting $c$.  Figure \ref{fig:tshprofile} depicts the resulting pairwise relationship.  Rational function fitting in \textbf{Step 5} yields the following functional relationship:  $c = \frac{0.236}{k_{34}}$.  The identifiable combination is therefore $ck_{34}$.
\\
\\

\begin{figure*}
\centering
\includegraphics[width=0.32\textwidth]{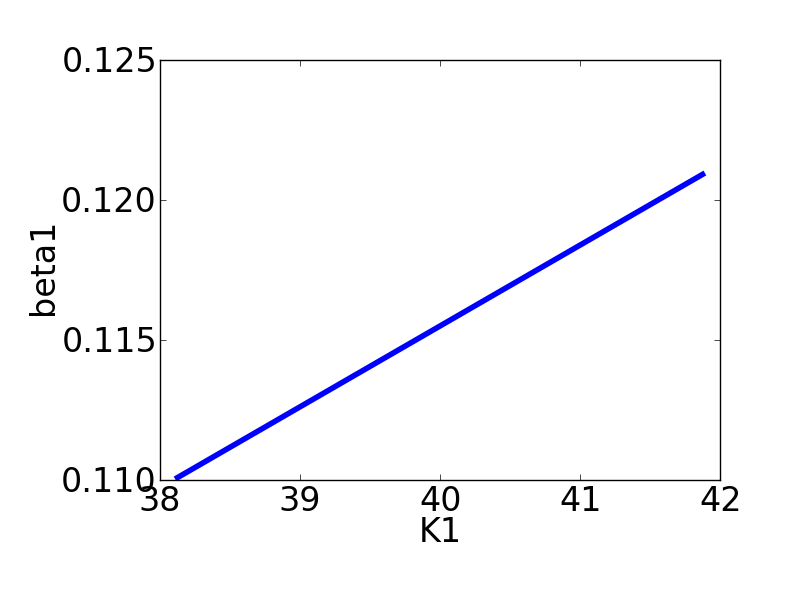}
\includegraphics[width=0.32\textwidth]{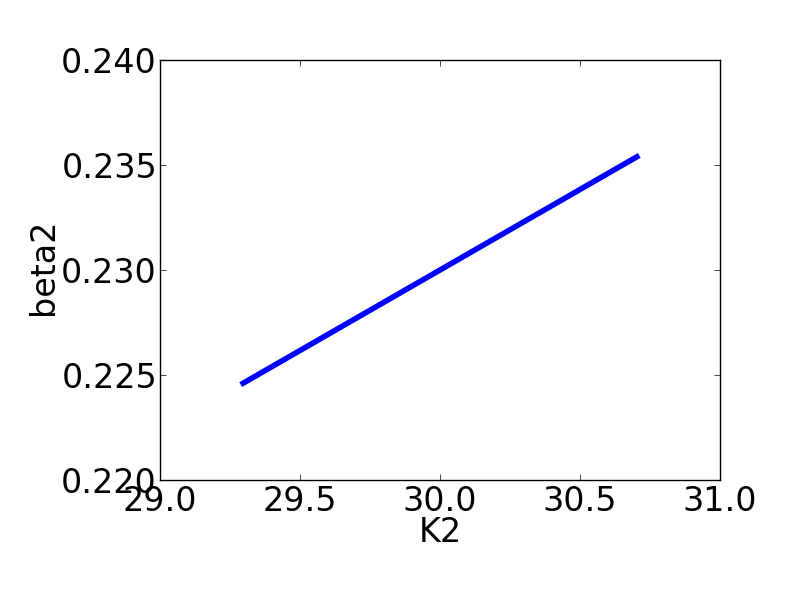}
\includegraphics[width=0.32\textwidth]{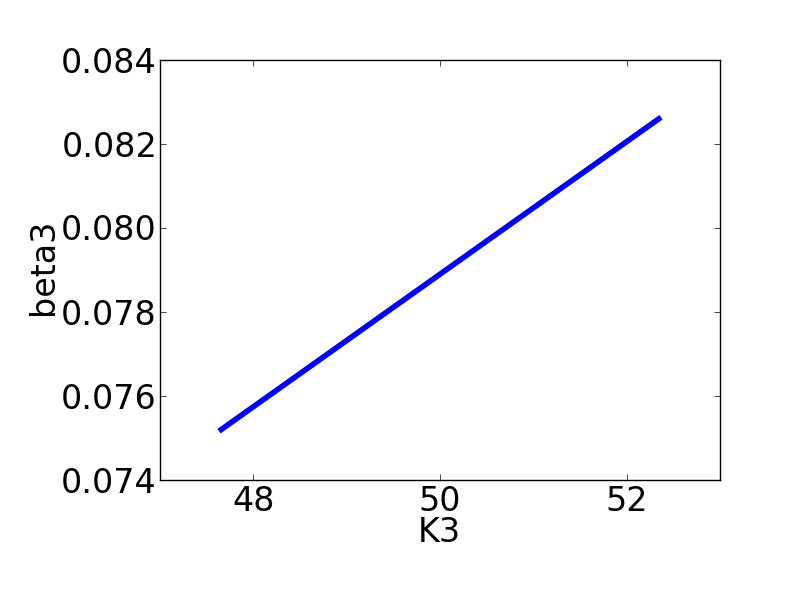}
\caption{Parameter relationships determined from the repressilator likelihood profile in Example 5.}
\label{fig:represprofile}
\end{figure*}

\noindent \textbf{Example 5: Repressilator Model}.  The repressilator model of Elowitz and Leibler \cite{Elowitz2008} is a nonlinear model of a synthetic cellular oscillator, generated from a cycle of three inhibitory circuits. This model has six ODEs, three output equations, and 19 parameters, including a Hill function with a fitted coefficient $n$, making the usual differential algebra approach inapplicable to this model as well. The model equations are given by:
\begin{equation}
\begin{aligned}
    \dot{m_i} &= \alpha_{0i} + \frac{\alpha_i}{1 + \left(\frac{p_{i-1}}{K_{i-1}}\right) ^n} - k_{degmi}m_i \\
    \dot{p_i} &= \beta_i m_i - k_{degpi} p_i \\
    y_i &= m_i
\end{aligned}
\label{eq:repres}
\end{equation}
for $i = 1,2,3$ modulo 3, where $m_i$ is the concentration of mRNA and $p_i$ is the concentration of protein. We suppose that gene expression data is taken, so that $y_i$ measures the concentration of each of the mRNA species. The model initial conditions were taken to be $m_1(0) = 1$ and all other variables were zero at $t=0$. The parameter descriptions and values are given in Table \ref{tab:represparams}. We constructed the sensitivity matrix in \textbf{Step 1} using the concatenation of all three output time series $\mathbf{y} = [y_1(t_1), \dots, y_1(t_n),y_2(t_1), \dots,y_2(t_n),y_3(t_1), \dots,y_3(t_n)]$. The resulting FIM computed has rank 16, indicating the presence of identifiable combinations.  All parameters except the three $K_{i-1}$'s and three $\beta_i$'s are identifiable, with \%CV's $>3000$ for these two and $< 100$ for the remaining parameters.  In \textbf{Step 2}, all parameters have individual \%CV's $<100$, so none are insensitive. The subset search in \textbf{Step 3} returns three pairs:  $(\beta_1, K_{1})$, $(\beta_2, K_{2})$, and $(\beta_3, K_{3})$.  In \textbf{Step 4}, we profile each $\beta_i$, fitting the corresponding $K_i$ (shown in Figure \ref{fig:represprofile}).  Per \textbf{Step 5}, we fit rational functions to the pairwise parameter plots, resulting in the following three functions:
\begin{equation}
\begin{aligned}
    K_{1} &= \frac{284.74\beta_1}{0.822} \\
    K_{2} &= \frac{160.338\beta_2}{1.229} \\
    K_{3} &= \frac{760.082\beta_3}{1.199}
\end{aligned}
\label{eq:represprofile}
\end{equation}
yielding identifiable combinations $K_{1}/\beta_1$, $K_{2}/\beta_2$, and $K_{3}/\beta_3$.

\begin{table} \centering \small 
\begin{tabular}{ccl}
\textbf{Parameter} & \textbf{Values ($i=1,2,3$)} & \textbf{Description} \\
\hline
$\alpha_{0i}$ & $5 \times 10^{-4}, 1 \times 10^{-4}, 9 \times 10^{-4}$ & Basal mRNA \\
 &  & transcription rate\\
$\alpha_i$ & $0.5$, $0.7$, $0.8$ & Regulated mRNA \\
 &  & transcription rate\\
$\beta_i$ & $0.1155$, $0.23$, $0.0789$ & Translation rate \\
$k_{degmi}$ & $0.005776$, $0.00987$, $0.00345$ & mRNA \\
 &  & degradation rate\\
$k_{degpi}$ & $0.001155$, $0.00059$, $0.004982$ & Protein\\ 
 &  & degradation rate\\
$K_{i}$ & $40$, $30$, $50$ & Inhibition constant \\
$n$ & 2 & Hill coefficient
\end{tabular}
\caption{Parameter values and descriptions for the repressilator model in Example 5. }
\label{tab:represparams}
\end{table}

\section{Conclusions} \label{sec:disc}

In this paper, we have demonstrated a numerical approach for structural identifiability analysis.  This approach extends the profile likelihood procedure developed by Raue \cite{Raue2009} with a focus on identifiable combinations. We perform a rank search on FIMs generated from parameter subsets to precondition the profile likelihood and generate functional forms for the identifiable combinations. The functional forms of the identifiable combinations can be used in several ways, for example by reparameterizing model in terms of these combinations, or by calculating the FIM-based variances for the identifiable combinations and using these as the estimated quantities in the model \cite{Barrett1998}. 

The rank search procedure detailed in \textbf{Step 3} of Section \ref{sec:numapproach} is similar in character to the sensitivity matrix-based method developed by Cintr\'{o}n-Arias \cite{CintronArias2009}, although in this case applied to the problem of identifiable combinations.  We focus on identifying parameter subsets for which profile calculations will uncover pair-wise functional relationships between unidentifiable parameters.  In particular, Examples 2 and 3 exemplify situations where pre-conditioning is necessary to avoid excess degrees of freedom. The FIM can be computed quickly for large models and is a natural extension of the sensitivity matrix. Like the optimal transformation method \cite{Hengl2007}, this approach does not require a priori information regarding parameter combinations. However, the FIM-based pre-conditioning steps do not require multiple parameter estimations or optimization. Compared to analytical approaches, identifiability analysis using our procedure is feasible in a reasonable amount of time on common hardware and is applicable to a wide range of model structures.

A general hurdle for methods based on the likelihood profile approach of \cite{Raue2009, Hengl2007} is in combining the 
profile parameter relationships into combinations when there are more than two parameters involved in the combinations. In the examples given here, this was relatively easy to determine simply by inspection, but in general a more algorithmic approach to this problem would be useful. Additionally, many of the steps in this approach are naturally parallelizable, or can be simplified in an algorithmic way, for example the search process for finding nearly full rank subsets in \textbf{Step 3}. In our examples, we considered relatively small models with up to six ODEs and up to 19 parameters. The general method is applicable to models of any size, however implementing the method for sufficiently large and complex models will likely require additional attention to issues such as \%CV thresholds, linear algebra methods, integrator and optimization routines, etc.

This approach can also be extended to understanding practically identifiable combinations, where noisy data may induce dependencies or combinations between parameters in an otherwise structurally identifiable model (e.g. as in \cite{Eisenberg2013}). Additionally, this method can be implemented for a much broader class of models (ODE, delay differential equations, discrete models, etc.) for which the FIM and likelihood profiles are appropriate. However, as with many numerical approaches to identifiability, this method is local (i.e. examines a point or local neighborhood in parameter space). Nonetheless, since structural identifiability is often taken as a generic property \cite{Audoly2001}, this can be addressed by testing the model identifiability for a range of randomly chosen values for the parameters.

\bibliographystyle{unsrt}
\bibliography{\bibpath/ComputationalIdentifiabilityEisenbergHayashi.bib}

\end{document}